\documentclass[conference]{sig-alternate-10pt}

\usepackage{array}
\usepackage{amssymb,epsfig,latexsym,epsf,color}
\usepackage{algorithm, algorithmicx, algpseudocode}
\usepackage{graphicx,amsmath,amssymb,latexsym}
\usepackage{subfigure}
\usepackage{paralist}
\usepackage{bm}
\usepackage{times}
\usepackage{multirow}
\usepackage{pseudocode}
\usepackage{balance}
\usepackage{color}
\usepackage{cite}

\usepackage[footnotesize]{caption}

\usepackage{enumitem}
\setitemize{noitemsep,topsep=0pt,parsep=0pt,partopsep=0pt}

\def\S{\mathcal S}

\newcommand{\beq}{\begin{eqnarray*}}
\newcommand{\eeq}{\end{eqnarray*}}
\newcommand{\beqn}{\begin{eqnarray}}
\newcommand{\eeqn}{\end{eqnarray}}

\DeclareFixedFont{\BX}{U}{cmss}{bx}{n}{10}

\begin{document}

\title{Making 802.11 DCF Optimal: \\Design, Implementation, and Evaluation}

\numberofauthors{3}

\author{
\alignauthor
Jinsung Lee, \\Yung Yi, Song Chong\\
       \affaddr{EE Dept., KAIST}\\
       \affaddr{Daejeon, Korea}\\
%       \email{ljs@netsys.kaist.ac.kr, \\
%        \{yiyung,songchong\}@kaist.edu}
\alignauthor
Bruno Nardelli, \\Edward W. Knightly\\
       \affaddr{ECE Dept., Rice Univ.}\\
       \affaddr{Houston, TX, USA}\\
%       \email{\{bn4, knightly\}@rice.edu}
\alignauthor
Mung Chiang\\
       \affaddr{EE Dept. Princeton Univ.}\\
       \affaddr{Princeton, NJ, USA}\\
%       \email{chiangm@princeton.edu}
}
%
%\author{\IEEEauthorblockN{Jinsung Lee,\\Yung Yi and Song Chong}
%\IEEEauthorblockA{EE Dept., KAIST\\Daejeon, Korea\\
%Email: ljs@netsys.kaist.ac.kr, \\\{yiyung,songchong\}@kaist.edu}
%\and
%\IEEEauthorblockN{Bruno Nardelli\\and Edward W. Knightly}
%\IEEEauthorblockA{ECE Dept., Rice Univ.\\Houston, TX, USA\\
%Email: \{bruno.nardelli, knightly\}@rice.edu}
%\and
%\IEEEauthorblockN{Mung Chiang}
%\IEEEauthorblockA{EE Dept. Princeton Univ.\\
%Princeton, NJ, USA\\
%Email: chiangm@princeton.edu}}

\maketitle

\begin{abstract}
This paper proposes a new protocol called {\em Optimal DCF} (O-DCF). Inspired by a sequence of analytic results, O-DCF modifies the rule of adapting CSMA parameters, such as backoff time and transmission length, based on a function of the demand-supply differential of link capacity captured by the local queue length. Unlike clean-slate design, O-DCF is fully compatible with 802.11 hardware, so that it can be easily implemented only with a simple device driver update. Through extensive simulations and real experiments with a 16-node wireless network testbed, we evaluate the performance of O-DCF and show that it achieves near-optimality, and outperforms other competitive ones, such as 802.11 DCF, optimal CSMA, and DiffQ in a wide range of scenarios.
\end{abstract}

%\begin{keywords}
%Random access, CSMA, 802.11 implementation, experiment
%\end{keywords}

\section{Introduction} \label{sec:intro}

% 802.11 DCF is heavily used in

%\note{1. It is quite well-known that 802.11 DCF has a lot of problems...}

Extensive research documents inefficiency and unfairness of the standard
802.11 DCF and suggests ways to improve it. Some proposals take a
clean-slate approach to redesign CSMA. Optimality in performance can
sometimes be proved under idealized assumptions such as no collision or 
perfect synchronization. Other proposals are constrained by operating over
legacy 802.11 hardware with only a device driver update, but are often
unable to attain optimality. In this paper, we propose a new protocol,
called {\em Optimal DCF} (O-DCF), which demonstrates that a collection
of design ideas can be effectively combined to transform legacy 802.11
DCF into a near-optimal yet also practical protocol.

%There is no shortage of research results documenting inefficiency and
%unfairness of the standard 802.11 DCF, nor is there a shortage of ideas
%to improve it. Some of the proposals in the vast literature take a
%clean-slate approach to redesign CSMA. Optimality in performance can
%sometimes be proved in the ideal cases such as no collisions or perfect
%synchronization. Other proposals are constrained by operating over the
%legacy 802.11 hardware with only a device driver update, but is often
%unable to attain optimality. In this paper, we propose a new protocol,
%called {\em Optimal DCF} (O-DCF), which demonstrates that a collection
%of design ideas can be effectively combined to upgrade the legacy 802.11
%DCF into a near-optimal, practical protocol.

Among prior methods to improve 802.11 DCF is the seemingly
conflicting pair of random access philosophies: in face of collisions,
should transmitters become more aggressive given that the supply of
service rate may become lower than the demand (as in the recently
developed theory of Optimal CSMA (oCSMA), e.g., \cite{LJ08,SRS09,LYP10,qcsma,ME08})?
Or should they become less aggressive given that collisions signal a contentious RF
environment (as in a typical exponential backoff in today's 802.11 DCF)? We show 
that these two conflicting approaches are in fact complementary. The best combination
depends on the logical contention topology, but can be learned without
knowing the topology.

%Among this collection of ideas to improve 802.11 DCF is the seemingly
%conflicting pair of random access philosophies: in face of collision,
%should each transmitter become more aggressive given that the supply of
%service rate may become lower than the demand (as in the recently
%developed theory of Optimal CSMA (oCSMA), e.g., \cite{LJ08,SRS09,LYP10,qcsma,ME08}),
%or should it be more polite given that collisions signal a contentious RF
%environment (as in a typical exponential backoff)? We show that these
%two conflicting approaches are complementary, where the best combination
%depends on the logical contention topology, but can be learned without
%knowing the topology.

As developed in theory such as oCSMA, a product of access probability,
which is determined by contention window (CW) in the backoff,
and transmission length should be proportional to the supply-demand
differential for long-term throughput fairness. Towards the goal of achieving
high performance in practice, a good combination of access probability and
transmission length is taken, where such a good access probability is
``searched'' by Binary Exponential Backoff (BEB) in a fully distributed manner
to adapt to the contention levels in the neighborhood, and then transmission
length is suitably selected for long-term throughput fairness.  Thus, BEB is
exploited not just to conservatively respond to temporal collisions, as in 
standard 802.11, but also to be adapted to appropriate access aggressiveness
for high long-term fairness and throughput by being coupled with the transmission
length.
% However, in practice, access probability should be carefully chosen in a fully distributed manner
% depending on

% O-DCF is designed to inherit all potentials of oCSMA in theory and at the same
% time, efficiently bridge the gap between theory and practice by making it work
% on top of the legacy 802.11 hardware. The main issues that do not appear in
% theory but significantly impact the performance in practice are related to
% the ``good'' choice of contention window (CW) size and transmission length.
% Particularly, the CW size directly affects collision and short-term fairness,
% depending on contention pattern, with which transmission length is controlled to
% achieve optimality.

We first summarize three key design ideas of O-DCF:
\begin{compactenum}[\bf \em D1.]
\item Link access aggressiveness is controlled by both CW
  size and transmission length, based on per-neighbor local queue length
  at MAC layer, where the queue length quantifies supply-demand differential.
  Links with bigger differential (i.e., more queue buildup) are
  prioritized in media access by decreasing CW size and/or increasing
  transmission length.

\item The CW size and transmission length are adapted in a fully distributed
  manner, depending on network topology affecting contention patterns in
  the neighborhood. Each link chooses the initial CW size as a decreasing
  function of the local queue length and then increases the CW size against
  collisions (i.e., backoff).
  Transmission length is set proportional to some product of the CW size
  (at which transmission succeeds) and the local queue length.

\item When wireless channels are heterogeneous across links, e.g., a
  link with 2Mbps and another link with 6Mbps, the link capacity information is
  reflected in controlling access aggressiveness by scaling the queue
  length proportionally to the link capacity. This adaptive control
  based on link capacity ensures better fairness and higher
  throughput, since links with better channel condition are scheduled
  more than those with poor channel condition.

% For more efficient rate allocation, link capacity informat% ion is
  % used in controlling access aggressiveness by scaling the queue length
  % proportionally to varying channel condition among links with different
  % capacities, which often happens in practice since the quality of the
  % received signals changes dynamically by e.g., the relative locations
  % and mobility of nodes, which we call {\em channel heterogeneity}.
%\item Channel heterogeneity is reflected in controlling aggressiveness
%  by scaling the queue length proportionally to channel capacity that is
%  often time-varying, which thus yields more efficient rate allocation
%  among links with heterogeneous capacities.
\end{compactenum}
\smallskip

In {\bf \em D2}, for the case when nodes can sense each other and
contends symmetrically, the CW size is appropriately chosen to be a
reasonably low value to reduce collisions, and then the transmission
length is chosen to be a function of queue length. For the topology
called flow-in-the-middle (FIM)\footnote{Throughout this paper, we use
  `flow' and `link' interchangeably.} where inner and outer links have
different contention degrees (i.e., asymmetric contention), the CW size
of a link that experiences more contention is adjusted to be smaller
than those of other links with less contention, so that it can get enough
transmission chances and thus fairness is ensured.  We will show that this
selective control of CSMA parameters works well even in challenging
topologies in which 802.11 DCF yields severe performance degradation, such
as hidden terminal (HT), information asymmetry (IA), FIM, and packet
capture.

The key design ideas mentioned above are implemented through the
following protocol mechanisms:

\smallskip
\begin{compactenum}[\bf \em P1.]
\item Each transmitter maintains two queues for each neighbor,
  referred to as Control Queue (CQ) and Media Access Queue (MAQ).
  CQ buffers the packets from the upper-layer which are to be dequeued into MAQ.
  The size of MAQ refers to the local queue length in {\bf \em D2}, determining
  access aggressiveness by adjusting CW size and transmission length.
  The dequeue rate from CQ to MAQ is appropriately controlled to ensure
  (proportional) fairness and high throughput.

\item Once the initial CW size is chosen as a function of the size of MAQ,
  BEB ``searches'' for the CW size
  at which transmission becomes successful in a fully distributed
  manner. This successful CW size is used to choose the transmission
  length as described in {\bf \em D2}.

\item We adapt transmission length based on time rather than bytes to
  achieve time fairness under heterogeneous channels.
  %in the scenarios with asymmetric channel capacities, as suggested by other research, e.g.,\cite{idlesense}.
  To that end, we exploit information from the rate-adaptation module
  in the 802.11 driver to determine the proper number of bytes to send,
  according to modulation and coding rate in use.
\end{compactenum}
\smallskip

All of the above mechanisms can be implemented using unmodified %over the existing
802.11 chips, as we have done in evaluating O-DCF over a 16-node wireless
testbed. In particular, the mechanisms satisfy the following constraints of staying
within 802.11:

% Our O-DCF is proposed with the constraints that it does not necessitate
% a clean-slate design and should work well over the legacy 802.11
% hardware, which is crucial towards practicality. Typical 802.11 chipsets
% have the following features that cannot be nullified or controlled at
% the device driver.

\smallskip
\begin{compactenum}[\bf \em C1.]
\item {\em Interface queue (IQ).\label{const1}} 802.11 hardware has
  a FIFO queue for storing packets ready for actual transmission to the media
  such that neighbor-specific packet control such as CQ and MAQ necessitates
  additional queues on top of IQ.

\item {\em CW granularity.\label{const2}} CW values are allowed by
  only some powers of two and thus we can only choose the value from the set
  $\{2^n-1, n=1, \cdots, 10\}$.

\item {\em BEB.\label{const3}} BEB (in which CW doubles for each
  collision up to a maximum value) is often hard-coded and cannot be nullified
  by software control. Also, once BEB is started, a typical device driver, e.g., 
  MadWiFi \cite{madwifi}, does not allow the driver to read the CW size at
  which the transmission is successful.

\item {\em Maximum aggregate frame size.\label{const4}} The packet size
  is bounded by a value that depends on the 802.11 chipset. For chipsets
  supporting packet aggregation, the packet size is constrained by a maximum
  aggregation size, e.g., 64 KB in 802.11n \cite{80211n}. Otherwise, it is
  constrained by a much shorter size, e.g., 2304 B in 802.11a/b/g.
  %Thus, we need to devise a method beyond a maximum packet size, e.g., 2304 B in 802.11a/b/g \cite{80211}.

%\item {\em Maximum aggregation size.} There is a constraint on the
%  maximum aggregation size to prevent long channel monopolization by
%  some node, e.g., 64 KB in 802.11n \cite{80211n}. \note{YY: what about your
%  implementation based on AIFS etc. Still, there is a limitation?}
\end{compactenum}
\smallskip

To evaluate the performance of O-DCF, we have implemented O-DCF on a wireless
testbed with 16 nodes as well as simulator for large-scale scenarios
that are difficult to be configured in the real testbed.  By comparing
O-DCF with 802.11 DCF, two versions of oCSMA, and DiffQ \cite{ASSI09},
we observe that in presence of conditions that are known to be critical
to other CSMA protocols,
% producing severe performance degradation and large throughput disparities,
O-DCF achieves near-optimal throughput, fairly distributed among flows,
with up to 87.1\% fairness gain over 802.11 DCF while keeping high network
utility in our experiments.

\section{Related Work}
\label{sec:related}

% There exists vast research literature in the design of wireless protocols for
% throughput and utility maximization (for a complete survey, see \cite{snum, LSS06}
% and the references therein).

% Starting from the seminal work on Max-Weight scheduling \cite{TE92}, some studies
% have proposed centralized algorithms that achieve maximum stability, i.e., support any
% set of traffic arrival rates for which there exists a scheduler that stabilizes
% queues within the network. Despite the importance of such works, centralized
% scheduling is hard to implement and introduces high control overhead \cite{YPC08}.
% Thus, distributed random-access protocols such as CSMA remain the preferred
% choice for most practical applications.
%\note{Q: Since our story is about enhancing 802.11 CSMA, do we need to add
%  more references on the efforts on enhancing 802.11 over multi-hop
%  networks out of some of zillions of papers
%in literatures?}

\noindent
{\bf \em Problems and Enhancements of 802.11.}
%\note{YY: You need to classify the existing approaches into some
%  categories, because we have to claim that they are not necessarily
%  wrong, but we take a different approach. I want to say that our
%  approach is often an orthogonal effort. For example, refer to the
%  recent mobicom papers that I mailed to you with Jeongseul. Roy at
%  duke's approach. They are actually compatible with ours right?}
Numerous papers have reported the performance problems of 802.11 DCF,
and proposed many solutions to them.  To name just a few, 802.11 DCF has
severe performance degradation and throughput disparities among
contending flows in the topologies such as HT, IA, FIM, and packet
capture \cite{bhademshezha94, MTE06, XK05, STYH09}, and heterogeneous
link capacities \cite{anomaly03}. We classify the solution proposals
into the efforts of MAC and PHY layers.  Some papers proposed
new access methods such as dynamic adjustment of CW under 802.11
DCF, e.g., \cite{tuning, idlesense}
%\cite{tuning, idlesense}
and there exist the implementation researches along this line, e.g., \cite{SS06, GHRD07}.
Other work presented efficient aggregation schemes and their real
implementations for improving throughput, e.g., \cite{aggregation, hydra08}.
%\note{distributed access methods? aggregation? sounds too general}
Note that most implementations mentioned above individually focused
on some specific topologies such as fully-connected (FC) case, and do not
explicitly consider problematic ones such as HT, IA, and FIM.
With the aid of PHY-layer, there are totally novel approaches, e.g., \cite{csmacn,
  crma11}. This kind of work exploits more information from PHY layer
and/or applies new PHY technologies other than CSMA. In O-DCF, we extend
MAC-level approach by controlling the CW size as well as the
transmission length based on the demand-supply differential, thus
yielding more performance benefit over 802.11 DCF. The PHY-layer based
approach is somewhat orthogonal to our approach, which even can be
integrated with O-DCF for further performance improvement.

\smallskip
\noindent{\bf \em Optimal CSMA.}
Recently, analytical studies proved that, under certain assumptions,
queue-length based scheduling via CSMA can achieve maximum throughput
without message passing e.g., \cite{LJ08,SRS09,LYP10}, which is referred
to as oCSMA in this paper.  Furthermore, multiple theoretical work
presented solutions based on the similar mathematical framework, each of
them focusing on different aspects of the protocol operation
\cite{qcsma,ME08,AYTM10,TM11}.  Our work is in part motivated by oCSMA theory,
but as reported in \cite{JJYSAM09,bjkysem11} and our evaluation, there
still exist many gaps between oCSMA theory and 802.11 practice.

\smallskip
\noindent{\bf {\em Implementation of queue based CSMA.}}
A limited number of work on the implementation of oCSMA exists, mainly
with focus on evaluation \cite{JJYSAM09,bjkysem11}.  They show that
multiple adverse factors of practical occurrence not captured by
the assumptions behind the theory can hinder the operation of oCSMA, introducing severe performance
degradation in some cases \cite{bjkysem11, JHYS12}. Other work is
devoted to bridge the gaps between theory and practice by
reflecting queue length over 802.11 \cite{ASSI09,ADPA09,xpress}.  In
\cite{ASSI09}, the authors implemented a heuristic differential backlog
algorithm (DiffQ) over 802.11e.  EZ-Flow was proposed to solve
instability due to large queue build-ups in 802.11 mesh networks
\cite{ADPA09}. Very recently, the authors in \cite{xpress} implemented a
backpressure scheduling based on TDMA MAC, which operates in a
centralized way.

% Moreover, some of the
% first experimental evaluations of oCSMA have shown that
%In particular, they show that the performance of oCSMA can be highly penalized in the presence of
%hidden terminals, asymmetric interference and asymmetric channel capacities,
%as well as upper-layer congestion-control mechanisms.
% On the other hand,
%Finally, \cite{TJRN11} derives a distributed protocol that maximizes throughput
%in the presence of hidden terminals, assuming all network nodes are synchronized.

% In contrast with the aforementioned work, we present the first asynchronous
% CSMA based protocol designed to work on top of the legacy 802.11 hardware
% and attain high performance in the environments comprising hidden terminals,
% asymmetric interference and asymmetric channel capacities.
% In addition, O-DCF is designed to inherit all of the high potentials of oCSMA
% in theory, yet efficiently bridge the gap between theory and practice through
% a fully operational protocol implementation over real wireless hardware.
%by making it work on top of real wireless hardware.
%In addition, different from most existing work, we experimentally validate our design with a fully
%operational protocol implementation over real wireless hardware.

%%% Local Variables:
%%% mode: latex
%%% TeX-master: "main"
%%% End:

\section{How Does O-DCF Work?}\label{sec:eocsma}

%\subsection{Overview}
We start by listing four key elements in O-DCF.

\begin{compactenum}

  \item {\em Section~\ref{subsubsec:queue_structure}.}
    Each transmitter is equipped with two queues for each neighbor,
    CQ and MAQ. CQ is a buffer to store the packets from the
    upper-layer and its dequeue rate into MAQ is controlled by
    a certain rule in strict relation to proportional fairness.

    % To achieve optimality in throughput and fairness, each link's contention
    % aggressiveness is adjusted by CW size and transmission length, based
    % on the historical queue differential between traffic load and effective
    % capacity over the link in the network (the mechanism is in Section \ref{subsubsec:queue_structure}).
    % unsuitable for symmetric, high-collision scenarios, where decreasing
    % CW over multiple flows worsens the problem. In our protocol,
    % we adapt aggressiveness based on transmission length to avoid too much
    % CW reductions.

  \item {\em Section~\ref{subsubsec:cw_choice}.}
    The size of MAQ, which quantifies the
    differential between demand and supply, is used to control the link access
    aggressiveness by adjusting CW size and transmission length. First, the initial CW size
    is determined by a sigmoid function of the size of MAQ, and then BEB is applied
    for collisions. Thus, the link is prioritized when the demand-supply differential
    becomes large.

    % In the transmission-length based aggressiveness adaptation,
    % transmitter's CWs should be carefully chosen. Large CWs reduce
    % collisions, but introduce large inter-transmission times. O-DCF
    % employs BEB to keep inter-transmission times low in low-collision
    % scenarios, while CWs increase to reduce collision probabilities only
    % for symmetric, high-collision scenarios.
  \item {\em Section~\ref{subsubsec:trans_choice}.}  The transmission
    length of each link is determined by the product of the success CW
    size, i.e., the CW size at which transmission succeeds, and MAQ
    size.  Typically, the success CW size is often hard to know from the
    device driver, thus we employ a method of estimating the success CW size.

    % We adapt the CW size depending on the topological features without any
    % message passing. Specifically, for symmetric contention scenarios, all nodes
    % increase CW values so that all nodes can succeed in their transmissions,
    % while for asymmetric contention scenarios, nodes have different CW values
    % according to their individual contention levels (the mechanism is in Section \ref{subsubsec:cw_choice}).

    % We control the transmission length in conjunction with CW adaptation,
    % which prevents severe performance degradation from too much CW reduction and
    % guarantees high throughput performance in case of highly contending scenarios
    % and even with nodes out of carrier-sense range.
    % (the mechanism is in Section \ref{subsubsec:trans_choice}).
    % While BEB works well for symmetric collisions, it does not
    % work very well for colliding nodes with asymmetric interference. It
    % just doubles the CW of the links that suffer from transmission losses.
    % In O-DCF, we adaptively set CW$_{\min}$ parameter of BEB based on a
    % decreasing function of the virtual queue length.  Such a function is
    % selected to span CW values in a sufficiently long range to attain high
    % performance with asymmetric interference, even with transmitters out
    % of carrier-sense range.
  \item {\em Section~\ref{subsubsec:channel}.}  Channel heterogeneity is
    reflected by scaling the MAQ size by the
    link capacity.  This gives more priority to the links with better
    channel conditions in media access, ensuring more efficient rate
    allocation in terms of time fairness.

    % We reflect channel heterogeneity in determining the contention aggressiveness
    % so that nodes with different link capacities can achieve more efficient rate allocation
    % in terms of time fairness (the mechanism is in Section \ref{subsubsec:channel}).
    % In O-DCF, we adapt transmission lengths based on time rather than
    % bytes to achieve proportional fairness in scenarios with asymmetric channel
    % capacities, as suggested by other research, e.g., \cite{idlesense}. To that
    % end, we exploit information from the rate-adaptation module in 802.11 drivers
    % to determine the proper number of bytes to send depending on the modulation rate in use.
  \end{compactenum}

%\subsection{Protocol Mechanisms}

\subsection{Supply-Demand Differential}\label{subsubsec:queue_structure}

%\note{Explain how virtual queue is maintained. When they are
%  updated... arrivals.. virtual arrival or real arrival... virtual queue
%is nothing but a scaled queue length...}

%\begin{equation}
%  \label{eq:vq_update}
%  q_l[t+1]=\bigg[q_l[t] + b\big(A_l[t] - S_l[t]\big)\bigg]^{q_{\max}}_{q_{\min}},
%\end{equation}

%\note{Explain $b$ from engineering's point of view, not just call it a
%  step size... $b$ is something that slows down queue variations...}
%
%\note{not sure whether we have to mention qmax and qmin... People will
%  be attacking on how to choose such parameters..}
%
%\note{What is $t$ in the real implementation? The word frame does not
%  make sense...}
%
%\note{May need to explain the connection between virtual queue and real
%  queue using a figure...}

\begin{figure}[t!]
  \begin{center}
    \includegraphics[width=0.85\columnwidth]{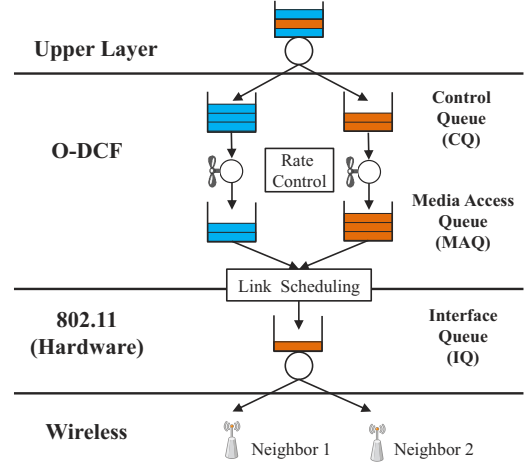}
    \smallskip
    \caption{Queue structure of O-DCF.}
    \label{fig:queue_structure}
  \end{center}
\end{figure}

%%%%%%%%%% Jinsung's writing
In 802.11 DCF, the packets from the upper-layer are enqueued to IQ
%so-called interface queue (IQ)
at the 802.11 chip for media access.  In O-DCF, we maintain two
per-neighbor queues (CQ and MAQ) over IQ (due to {\bf \em
  C\ref{const1}}), to balance the link's supply and demand through fair media
access, as shown in Figure~\ref{fig:queue_structure}.  Denote by
$Q^C_l(t)$ and $Q^M_l(t)$ the sizes of CQ and MAQ for each link $l$
at time $t.$ We further maintain a variable $q^M_l(t)$, which is simply
the scaled version of $Q^M_l(t)$, i.e., $q^M_l(t) = b Q^M_l(t),$ where
$b$ is some small value, say 0.01.  The value of $Q^M_l(t)$ is crucial
in O-DCF in that both the dequeue rate from CQ to MAQ and the
aggressiveness of media access tightly rely on $Q^M_l(t).$

First, we control the dequeue rate from CQ to MAQ (when CQ is
non-empty), such that it is inversely proportional to $q^M_l(t),$ by
$V/q^M_l(t),$ where $V$ is some constant controlling the sensitivity of
the dequeue rate to MAQ. Second, in terms of aggressiveness in media
access, an initial CW size and transmission length, which determine the
dequeue rate of MAQ, is set as a function of $q^M_l(t)$ (see
Sections~\ref{subsubsec:cw_choice} and \ref{subsubsec:trans_choice} for
details). % By transmission length, we mean that multiple packets can be
% transmitted back-to-back, just like packet aggregation in 802.11n.
Then, whenever a new arrival from CQ or a service (i.e., packet
transmission) from successful media
access occurs, $Q^M(t)$ is updated by:
\begin{multline}
  \label{eq:vq_update}
  Q^M_l(t+\delta t)=
  \bigg[Q^M_l(t) + \cr \big(\text{arrival from
    CQ}-\text{service from MAQ})\big)\bigg]^{Q_{\max}}_{Q_{\min}},
\end{multline}
where $\delta t$ is the elapsed time of the next arrival or service
event after $t.$ The service from MAQ occurs when the HOL (Head-Of-Line)
packet of MAQ is moved into IQ.
%\footnote{Thus, strictly speaking, a
%  service from MAQ is not an actual packet transmission due to {\bf \em C\ref{const1}}.
%  %This is unavoidable because we have the constraint of implementing O-DCF on
%  %top of 802.11 hardware.
%  To minimize the time gap between the service from MAQ and the actual
%  transmission, we reduce the buffer limit of IQ to one through a device
%  driver modification.}.
  For multiple neighbors, the largest MAQ is
served first; %if there are more than a packet to be served in that MAQ,
%multiple packets from the same MAQ can be scheduled until the chosen
%transmission length expires.
%\note{Here, we use multiple packets. What
%  happens to the word ``maximum packet size''?}
If the chosen transmission length exceeds the maximum aggregation size,
multiple packets from the same MAQ are scheduled in succession.
% If there are more than one neighboring link, the link who has the
% highest MAQ buildup is served first.  The time gap between this
% dequeueing process and actual transmission can be minimized by
% reducing the buffer size of IQ.
$Q_{\max}$ is the physical buffer limit of MAQ, but $Q_{\min}$ is set
by us as some positive value to prevent too severe oscillations of $Q^M_l(t),$ where
too small $Q_{\min}$ would sometimes lead to impractically high injection rate
from CQ to MAQ, when $Q^M_l(t)$ (and thus $q_l^M(t)$) approaches zero, where recall that the
injection rate is $V/q_l^M(t).$

%\subsubsection{$CW_{\min}$ with BEB}
\subsection{Initial CW with BEB}
\label{subsubsec:cw_choice}

%%%%%%%%%% Jinsung's writing

% \noindent{\bf \em $CW_{\min}$ with BEB.}
When a link $l$ is scheduled at time $t$, its initial CW size
($CW_l(t)$) and the number of bytes to transmit ($\mu(t)$) are set adaptively
as a function of the size of MAQ. First, $\mu(t)$-byte transmission over
$l$ is assigned with the following CW size:
%\note{YY: Do we need to have subscript $l$ for CWmin?}
\begin{eqnarray}
  \label{eq:initial_cw}
%  \tilde{p}_l[t] = f(q_l) = \frac{\exp(q_l[t])}{\exp(q_l[t]) + C},
    CW_l(t) = \frac{2\big(\exp(q_l^M(t))+C\big)}{\exp(q_l^M(t))}
    - 1,
\end{eqnarray}
where $C$ is some constant whose suitable value will be discussed later.
We want to use $CW_l(t)$ as the minimum CW ($CW_{\min}$) in 802.11 DCF, but
due to the constraint {\bf \em C\ref{const2}} that 802.11 hardware only allows CW values as
powers of two, we use one of
possible values closest to \eqref{eq:initial_cw} as $CW_{\min}.$ Then,
the media access is attempted after the time (in mini-slots) randomly
chosen from the interval $[0,CW_{\min}].$
Intuitively, we assign higher aggressiveness in media access for larger MAQ
size, remarking that \eqref{eq:initial_cw} is decreasing with $q_l^M(t).$
Whenever collision happens, this CW value exponentially
increases by BEB from {\bf \em C\ref{const3}}.
We will discuss later that BEB is not just an inevitable component due to
our design constraint of the 802.11 legacy hardware,
but is also an important component to improve performance inside our design rationale.

% Note that 802.11 hardware only allows CW
% values as powers of two, as described in {\bf \em C\ref{const2}}.
% %Thus, the actual $CW_l$ is set by the one out of possible values closest to the RHS of
% %\eqref{eq:initial_cw} and is used for the minimum CW ($CW_{\min}$) of 802.11.
% Thus, the initial minimum CW ($CW_{\min}$) of 802.11 is

It is often convenient to interpret $CW_l$ with its corresponding access
probability $p_l$ using the relation $p_l = 2/(CW_l + 1)$ \cite{tuning,
  idlesense}, where the initial $CW_l$ selection in
\eqref{eq:initial_cw} is regarded as the following {\em sigmoid} function:
\begin{eqnarray}
  \label{eq:initial_access}
%  \tilde{p}_l[t] = f(q_l) = \frac{\exp(q_l[t])}{\exp(q_l[t]) + C},
    p_l(t) = \frac{\exp(q_l^M(t))}{\exp(q_l^M(t))+C}.
\end{eqnarray}
We delay our discussion on why and how this sigmoidal type function helps
and what choice of $C$ is appropriate to Section~\ref{subsubsec:sigmoid}.

\subsection{Transmission Length Selection}
\label{subsubsec:trans_choice}

Our design on selecting transmission length $\mu_l(t)$ whenever the
media is grabbed by link $l$ is to set $\mu_l(t)$ as a function of the
success access probability (equivalently, the success CW size) and the
size of MAQ (i.e., $Q_l^M(t)$).  Again, by the success CW size, we mean
that the transmission becomes successful at that CW size, which is often
larger than the initial CW size due to BEB. The rationale to search for
the success CW size lies in the fact that it is the actual value used in
media access for successful transmission.  However, such a success CW
size is often hard to be read by the device driver ({\bf \em
  C\ref{const3}}).  Therefore, we estimate it based on the equation
\cite{transport02} which discloses the connection between the initial CW
($CW_l$), collision ratio ($p_c$) and the success access probability
after BEB (denoted by $\tilde{p}_l$), given by:
\begin{equation}
  \label{eq:beb}
  \tilde{p}_l = \frac{2 q(1-p_c^{m+1})}{(CW_l+1))(1-(2p_c)^{m+1})(1-p_c)+q(1-p_c^{m+1})},
\end{equation}
where $q=1-2p_c$ and $m$ is the maximum retransmission limit. The value
of $p_c$ can be computed for arbitrary topologies \cite{MTE06}, if nodes
have the complete knowledge of topology via message passing which is
often expensive.  To know $p_c$ without such message passing, we
approximate $p_c$ by measuring the packet collisions over the last $t'$
seconds (say one second) and averaging them.  The measurement on $p_c$
is based on counting the unacknowledged number of transmitted packets
during the interval.

Using the measured success access probability $\tilde{p}_l(t),$ in
O-DCF, the transmission length (in mini-slots) is chosen by:
\begin{eqnarray}
  \label{eq:trans_length}
\mu_l(t) &=& \min \left(\frac{\exp(q_l^M(t))}{\tilde{p}_l(t)}, \bar{\mu}
  \right),
\end{eqnarray}
where $\bar{\mu}$ is the maximum transmission length (e.g., 64 packets in our
setting, which is similar to the maximum aggregation size in 802.11n to
ensure the minimum short-term fairness\footnote{For measuring short-term fairness, we use
the inverse of the largest time difference between two consecutive packet transmissions
for a flow.}
and prevent channel monopolization by some node).
Then, we convert the transmission length in the unit
of mini-slots of the 802.11 chipset into that in bytes so as to compute
the number of packets for aggregate transmission as follows:
\begin{equation}\label{eq:mu_conversion}
  \mu_l \textrm{(bytes)} = \mu_l \textrm{(slots)} \times c_l \textrm{(Mb/s)} \times t_{\text{slot}} \textrm{($\mu$s/slot)},
\end{equation}
where $c_l$ is the link capacity and $t_{\text{slot}}$ is the duration of a
mini-slot ($9 \mu$s in 802.11a).
When only a part of $\mu_l$ bytes is transmitted due to packetization, we maintain
a {\em deficit counter} to store the remaining bytes of the transmission length that
will be used in the next transmission.

\subsection{Channel Heterogeneity and Imperfect Sensing}\label{subsubsec:channel}

In practice, wireless channels are heterogeneous across users as well as
often time-varying. In such environments, most 802.11 hardware exploits
multi-rate capability of the PHY layer to adapt their rate, e.g.,
SampleRate \cite{bicket}. However, it is known that 802.11 DCF is
incapable of utilizing this opportunistic feature, leading to the waste
of resource called {\em performance anomaly} \cite{anomaly03}. In other
words, 802.11 DCF provides the equal chances to the links (on average),
in which case the low-rate links would occupy more time than the
high-rate ones, so that the performance degrades. To provide fairness
focusing on time shares instead of rate shares, namely {\em time-fairness} \cite{timefairness},
we slightly modify our rules in selecting the initial access probability
as well as the transmission length by replacing $\exp (q_l^M(t))$ by $\exp
(c_l(t)q_l^M(t))$ in \eqref{eq:initial_cw} and \eqref{eq:trans_length},
where $c_l(t)$ is (relative) link capacity of link $l$ at time $t$, as
theoretically verified by \cite{AYTM10}.

For imperfect sensing cases such as HT and IA scenarios, we also propose to
use a virtual carrier sensing via RTS/CTS signaling, as suggested in literature.
In O-DCF, unlike in the standard 802.11a/b/g, RTS/CTS signaling is conducted only
for the first packet within the transmission length.

\section{Why Does O-DCF Work?}\label{sec:why_eocsma}

\subsection{oCSMA Theory}\label{subsec:ocsma}

O-DCF is in part motivated by the recent research results on queue-based MAC
scheduling in theory community, see e.g., a survey \cite{YC09-survey} and
in particular, the studies on oCSMA %on so-called Optimal CSMA (oCSMA)
\cite{LJ08,SRS09,LYP10,qcsma,ME08}. oCSMA is characterized as a CSMA
that has a specific rule of setting backoff time and transmission
length. They are slightly different in terms of the models and
conditions, e.g., discrete/continuous, synchronous/asynchronous, or
saturated/unsaturated traffic. However, the key idea is largely shared; the
queue maintains the demand-supply differential, and the access
aggressiveness is controlled by the queue length, which, in turn,
depends on the demand (arrival) and the supply (transmission success).
It has been proved that as long as $p_l(t) \times \mu_l(t) =
\exp(W(q_l(t))),$ where $W(\cdot)$ is a weight function, optimality in
terms of throughput or fairness is ensured.

As an example, in the saturated model with infinite backlog,
the optimality can be stated as:
the long-term rate $\bm{\gamma}^\star= (\gamma_l^\star: l \in \cal{L})$
($\cal L$ is the set of all links) is the solution of the following optimization problem:
\begin{align}
  \label{eq:network_utility}
  \max   \quad \Sigma_{l \in \cal{L}}U(\gamma_l), \quad
  \hbox{such that}  \quad \bm{\gamma} \in \Gamma,
\end{align}
where $\Gamma$ is the throughput region that is the set of all long-term
rates possible by any MAC algorithm. It is proved that $\bm{\gamma}^\star$
can be realizable by oCSMA.
Of particular interest is the logarithmic utility function (i.e.,
$U(\cdot)=\log(\cdot)$), in which case the source rate is controlled by
$V/q_l(t)$ for some positive constant $V$. This source rate
control, together with oCSMA, achieves {\em proportional fairness},
which effectively balances fairness and efficiency.

% To summarize, oCSMA theory provides a guideline on how to choose the
% access probability and transmission length for each link as a function
% of demand-supply differential towards optimal fairness. However, we stress
We highlight that O-DCF is not just a naive implementation of oCSMA,
because many assumptions in the oCSMA theory, e.g., no collisions in the
continuous time framework, symmetric sensing, perfect channel holding, etc. 
do not hold in practice. Furthermore, O-DCF is constrained, to be fully compatible 
with 802.11 chipsets. More importantly, in theory, any combination of $p_l$ and $\mu_l$ 
works if their product is $\exp(W(q_l(t))).$ However, we need a careful
combination of them for high performance in practice. All of these issues 
will be elaborated in the following sections.

Among other results, \cite{bjkysem11} shows that symmetrically high
collision scenarios can induce an excessive aggressiveness by oCSMA
flows, which in turn can lead to the complete throughput degradation
(e.g., HT).  In addition, asymmetric collisions can introduce large
disparities in the throughput attained by different flows (e.g.,
IA). While oCSMA can help with reducing the aggressiveness of advantaged
flows, the range of values in which the CW size is adapted is
insufficient to provide high performance gains when the transmitters are
out of carrier-sense range. Finally, oCSMA excessively prioritizes links
with low channel quality, due to queue based aggressiveness
control. This introduces severe inefficiency to the performance of
oCSMA.

\subsection{Tension between Symmetric and Asymmetric Contention}

\subsubsection{Topological Dependence}
\begin{figure}[t!]
  \centering
  \subfigure[FC]{\includegraphics[width=0.2\columnwidth]{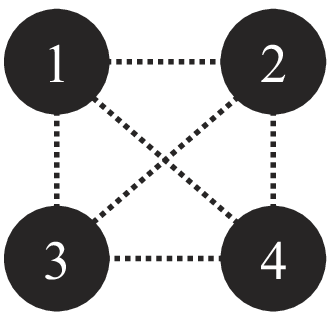}
    \label{fig:fc4}} \hspace{0.5in}
  %\subfigure[FIM with 3 outer flows]{\includegraphics[width=0.2\columnwidth]{figure/star4.eps}
  %  \label{fig:star4}}
  \subfigure[FIM]{\includegraphics[width=0.2\columnwidth]{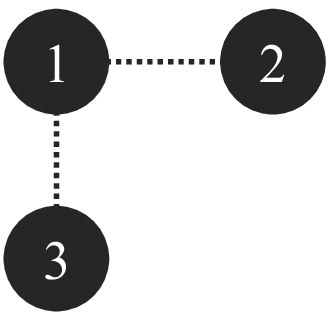}
    \label{fig:fim}}
  \caption{Example topologies' conflicting graphs. Vertices are the links in interference graphs;
    dotted lines represent interference; (a) FC with 4 links;
    (b) FIM with 2 outer links.}
  \label{fig:fc_star_topo}
\end{figure}

A good combination of two CSMA operational parameters for high
performance depends on contention topologies. O-DCF is designed to
autonomously choose the combination of access probability and
transmission length without explicit knowledge of topological
information. We provide our description, assuming that flows are
configured in either of the two ``extreme'' topologies: fully-connected (FC)
for symmetric contention and flow-in-the-middle (FIM) for asymmetric
contention (see Figure~\ref{fig:fc_star_topo}). As the name implies,
symmetry or asymmetry in contention refers to whether the contention level
is similar across flows or not. We use these two topologies just for ease
of explanation, and O-DCF generally works well beyond these two topologies.

Recall the two key design ideas of O-DCF: we first choose the
initial access probability as a sigmoid function of the queue length
and then let it experience BEB. To summarize, BEB is a key component in
symmetric contention (see Section~\ref{subsubsec:beb}) and the sigmoid
function based access probability selection is crucial in asymmetric
contention (see Section~\ref{subsubsec:sigmoid}), and both are important in
``mixture'' topologies (see Section~\ref{subsubsec:fim_fc}).

% differs for

% requires
% without topological information, we consider two extreme topologies:
% (i) fully connected (FC) topology for symmetric contention and (ii) FIM topology
% for asymmetric contention (see Figure~\ref{fig:fc_star_topo}).
% Obviously, theses topologies are insufficient to examine the full details of the
% required properties for high performance. Nonetheless, they can give us the key
% insights for optimal control of two parameters.

\subsubsection{O-DCF: How Does Exponential Backoff Help?}
\label{subsubsec:beb}

%Our approach for access probability selection is summarized as: (i)
%we first compute the {\em initial} access probability, $\tilde{p}_l[t],$
%as a function of the local queue length, where the function is designed
%to be a sigmoidal form, and (ii) BEB is run with the initial access probability.
%The access probability after BEB, denoted by $p_l[t],$ is regarded as the
%{\em actual} one in the sense that the transmission length $\mu_l[t]$ is
%computed by not with $\tilde{p}_l[t]$ but $p_l[t]$: $\mu_l[t] = \exp(q_l[t])/p_l[t].$

% the {\em actual} access probability $p_l(t)$ is
% finally found by the famous BEB. Note that the transmission length $\mu_l(t)$
% is set based on $p_l(t)$ following the equation $p_l(t) \times \mu_l(t) = \exp(q_l(t)).$

% BEB (Binary Exponential Backoff) with
% sigmoidal-type queue-based initial access probability selection.  We
% henceforth explain why this approach is practically appropriate to meet
% the requirements mentioned earlier. We first start by BEB, followed by
% how we choose the initial access probability selection. For clarity,
% denote by $p_l^{\text{I}}(t)$ and $p_l^{\text{B}}(t)$ the initial and
% BEB-experienced access probabilities, respectively.

% We first summarize the two key features of our proposed approach in
% selecting access probabilities, and then explain how they help with
% satisfying the requirements mentioned earlier.

%\medskip
%\noindent{\em (a) BEB (Binary Exponential Backoff)}
%\smallskip

%We first start by focusing on BEB, for a given initial access
%probability (discussed in (b)).

In symmetric contention, the access probabilities among the contending
flows should be reasonably low; otherwise, throughput will naturally
degrade.  Note that to guarantee fairness and high (long-term)
throughput, a tiny $p_l$ can work as it leads to almost no collision.
This is because in that case a significantly long transmission length would
recover the long-term throughput, as explained in oCSMA theory.
However, such a combination will experience a serious problem in short-term
fairness, where a maximum bound on transmission length to guarantee short-term
fairness is enforced in practice. For an automatic
adaptation to contention level, we utilize BEB as {\em a fully
  distributed search process} for the {\em largest} access probability (i.e.,
the smallest CW size) that lets the links access in presence of
collisions. This usage is in stark contrast to BEB in 802.11 DCF
that simply conservatively tries to avoid collisions.

% In the literature, it is known that BEB
% naturally wastes time resource and causes unfair channel access in
% short-time scale \cite{idlesense}.

% However, in O-DCF,
% we also vary the transmission length based on the resultant CW after BEB and the queue
% length, thus substantially reducing wasted time by BEB and guaranteeing short-term fairness
% among flows in symmetrically contending scenarios.

%The access probability of link $l$ which symmetrically contends with other links,
%such as FC topologies, should be {\em low} enough that throughput loss due to collisions
%is minimized and the aggravation cycle generated by the build-up of queues intensifying
%collisions does not happen.

%BEB is a mechanism that responds to packet collisions, which doubles the CW size
%per each collision. As an adaptation to contention level, we utilize BEB as a fully
%distributed search process for the ``best'' CW that lets the links access in presence
%of collisions. Note that BEB naturally wastes time resource, and also that in FIM-like
%topologies, as we confirmed in simulation and experiment, there are almost no collisions
%and BEB rarely operates. The selection of an initial access probability, as explained
%next, plays an important role of reducing the number of BEB in FC-like topologies as
%well as prioritizing the central flow in FIM-like topologies.

\subsubsection{O-DCF: Why Sigmoid Function?} \label{subsubsec:sigmoid}
%\medskip
%\noindent{\em Why sigmoidal?}

%\note{Q: Still, we have some logic for why $C=500,$ i.e., inflection
%  points.}

As opposed to symmetric contention, in asymmetric contention such as
FIM-like topologies, almost no collision occurs and thus BEB rarely operates
(we confirmed in Section~\ref{subsec:contention}). More importantly, in
this case, the starvation of the central flow is a major issue.
To tackle this, we require that the CW size (or the access probability)
of link $l$ that solely contends with many other links should be {\em small (or high)}
and thus {\em prioritized} enough that the link $l$ avoids rare channel access and
even starvation. %, as investigated in Section \ref{subsubsec:motive}.
To provide such {\em access differentiation}, we note that the flow in
the middle, say $l$, typically has a more queue buildup than the outer flows.
We denote the access probability $p_l$ by some function of queue length $q_l$,
i.e., $p_l=f(q_l)$.  Thus, it is natural to design $f(q_l)$ to be {\em
  increasing} for access differentiation.

The question is what form of the function $f(q_l)$ is appropriate
for high performance. To streamline the exposition, we proceed the discussion
with the access probability rather than the CW size. Toward efficient
access differentiation, we start by the $f$'s requirements: for any
link $l$ and for $q_{\min} \leq q \leq q_{\max},$

% We narrow down the good $f$ based on

% We henceforth explain that out of many increasing functions $f(\cdot),$
% a sigmoidal type function is an appropriate choice in practice.
% First, we have the following requirements for $f(q)$.
%\note{Jinsung, from here...}

\smallskip
\begin{compactenum}[\bf \em R1:]

\item $0 \leq f(q_l) \leq \bar{p} < 1,$ where $f(q_{\min}) \approx 0$ and
  $f(q_{\max})= \bar{p}.$ The largest access probability $\bar{p}$ should be
  strictly less than one to prevent channel monopolization.

\item $[f(q_l)]_2$ \footnote{We denote by $[x]_{2}$ the $1/2^i$ for some
    integer $i$, which is closest to $x,$ e.g., $[0.124]_2 = 1/2^3$ and $i=3$.}
    should span all the values in $\{1/2^i, i = 0, \ldots, 9\}$ each of
    which corresponds to the CW sizes $\{2^{i+1}-1, i = 0, \ldots, 9\}$ due to the
    CW granularity constraint {\bf \em C\ref{const2}}. 
    %Let us call $i$ the level of access probability or CW size.

% \item For a queue differential $\Delta q,$ $f(q_l + \Delta q) =
%   f(q_l) \times 2^{\Delta q}.$ This condition is regarded necessary for
%   efficient access differentiation as explained next.

\item The transmission lengths of the flows with heavy contention and
  those with light contention should be similar.

\end{compactenum}
\smallskip

The requirement {\bf \em R3} is important to prevent the central flow
from being starved. In asymmetric contention such as the
FIM-like topologies, the flows experiencing heavy contention such as
the central one in FIM has very rare chances to access the media. To guarantee
(proportional) fairness, it is necessary for such flows to select long
transmission lengths whenever holding the channel. However, as
mentioned earlier, the maximum transmission length should be bounded for
practical purpose such as short-term fairness. This implies that the
central flow often needs to stop the transmissions before its required
transmission length for optimal fairness is reached.
Efficient flow differentiation, which essentially prioritizes the flows
with heavy contention in terms of access probability, helps much by reducing
the required transmission length with a reasonable value (mostly shorter than 
the maximum transmission length) towards long-term fairness.
An extreme case for asymmetric contention is IA scenario where two flows have 
asymmetric interference relationship. We will explain in Section~\ref{subsec:sensing} 
that our flow differentiation helps a lot in providing (long-term) fairness 
in such a problematic scenario.

An intuitive way to realize flow differentiation is to set the access
probability of link $l$ to be $\exp(q_l)/K$ for some constant $K.$ Then,
the rule \eqref{eq:trans_length} enforces the transmission length to be
around $K,$ {\em irrespective of the contention levels of the flows}
(i.e., {\bf \em R3}).  However, to satisfy {\bf \em R1}, we use a slightly
different function that has a sigmoidal form, $f(q_l) =
\frac{\exp(q_l)}{\exp(q_l) +C},$ with some constant $C.$ This function
naturally makes the chosen access probability to be strictly less than one
for any $q_{\min} \leq q \leq q_{\max}$, unlike $\exp(q_l)/K$, because
it is increasing up to $f(q_{\max}).$ Clearly, the sigmoid function is
not exponential over the entire $q_{\min} \leq q_l \leq q_{\max}$
values. However, it suffices to have an exponential form up to $q_l'$
with $f(q_l') = 0.75,$ since for a larger $q_l > q'_l,$ the CW size approaches one
from the CW granularity (but, the access probability is set to be strictly less than one).
% (i.e., CW size 1) is chosen. \note{access probability should be strictly less than one...}

% One of the candidate functions satisfying {\em R1, R2,} and {\em R3} is
% a sigmoidal type function, $f(q_l) = \exp(q_l)/(\exp(q_l) +C),$ for some
% constant $C.$ Clearly, the sigmoidal function is not exponential over
% the entire $q_{\min} \leq q_l \leq q_{\max}$ values. However, it
% suffices to have an exponential form up to $q_l'$ with $f(q_l') = 0.75,$
% since for a larger $q_l > q'_l,$ from the CW granularity, the acces
% probability one (i.e., CW size 1) is chosen. Also, from {\em R1}, it
% would be more natural to goes to $1,$ as $q_l$ grows.

\begin{figure}[t!]
\centering
    \subfigure[Queue vs. $p$]{
        \includegraphics*[width=0.49\columnwidth]{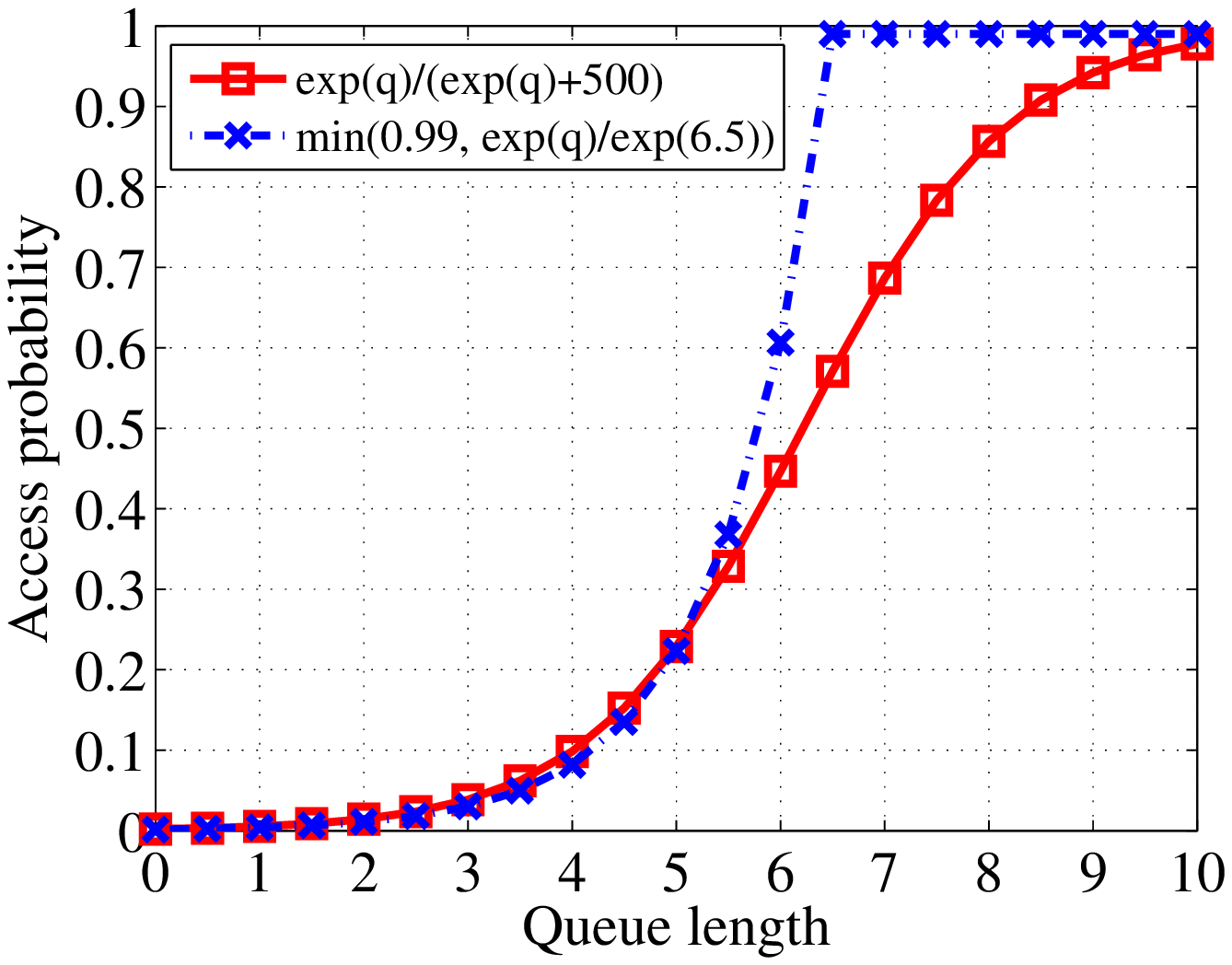}
        \label{fig:sigmoid_p}
    }
    \hspace{-0.35cm}
    \subfigure[Queue vs. CW]{
        \includegraphics*[width=0.49\columnwidth]{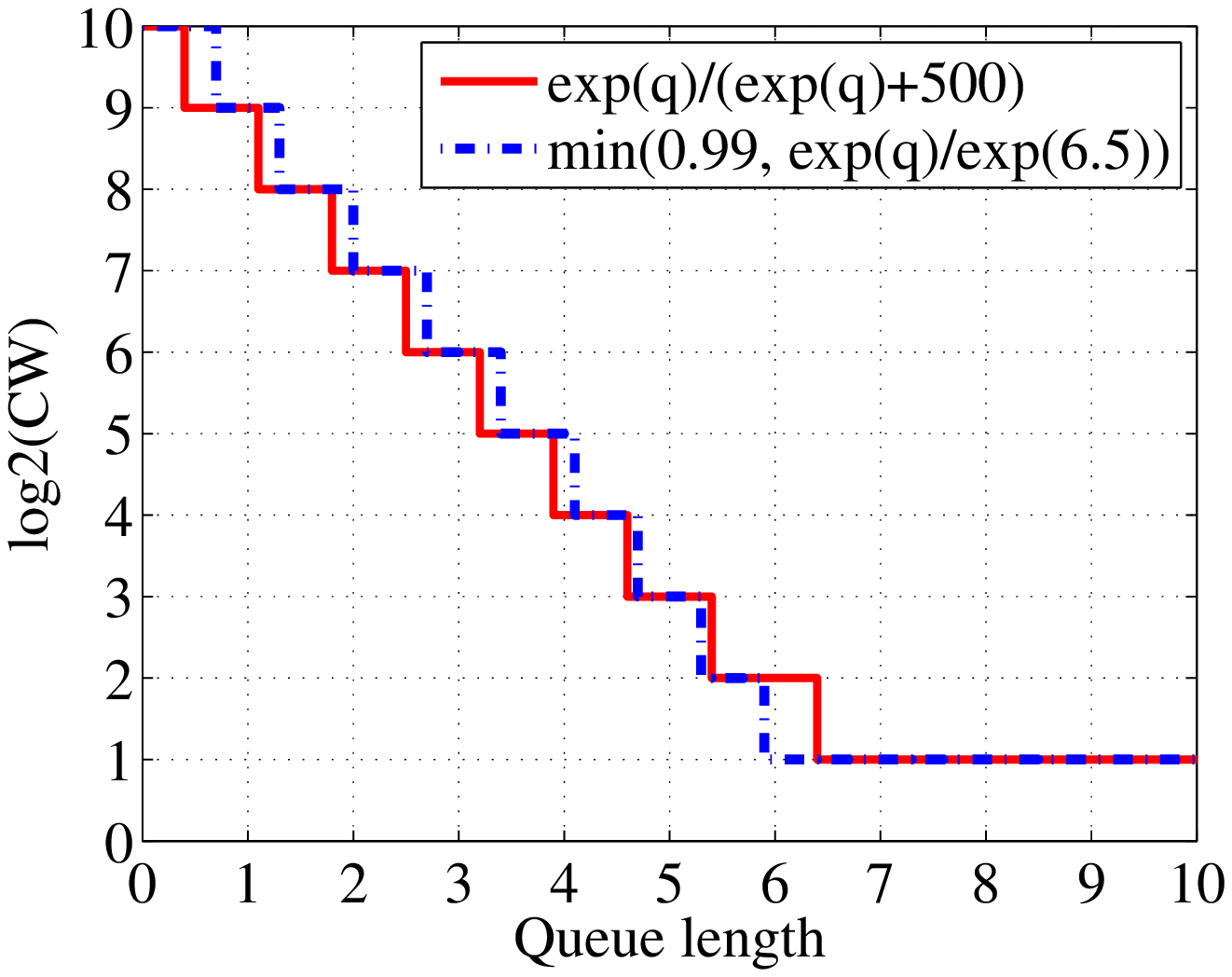}
        \label{fig:sigmoid_cw}
    }
    \caption{Illustration of sigmoid and exponential functions with respect to queue length.}
    \label{fig:sigmoid}
\end{figure}

Then, the next question is the inflection point, determined by the
constant $C.$ We choose $C$ around 500 due to the following reasons.  First,
for the resultant $p$ to span the whole feasible values (as in {\bf \em R2}),
$C$ should be greater than 500, i.e., $f(q_{\min})=
\frac{\exp(q_{\min})}{\exp(q_{\min})+500}\approx \frac{1}{512}$, where
$q_{\min}=bQ_{\min}=0.01$ (see Figure~\ref{fig:sigmoid_cw} in terms of
CW).  Second, the parameter $C$ determines the location of inflection
point in the sigmoid function. For example, let us consider the FIM
topology with four outer flows, where the central flow can have its
queue length of up to five, while the outer flows have much
smaller queue lengths usually less than two. To guarantee the
exponential increase in that curve, the $x$-axis at the inflection point
should be larger than five, implying that $C$ larger than 500 is
sufficient since $f(5) = \frac{\exp(5)}{\exp(5)+500} < \frac{1}{2}$.
However, too large $C$ values directly results in too long transmission
length from \eqref{eq:trans_length} because it yields a very low $p$.

Finally, two remarks are in order: (i) Due to the CW
granularity constraint {\bf \em C\ref{const2}}, the exponential function
$f(q_l) = \min (1,\exp(q_l)/\exp(6.5))$ leads to the similar CW
mapping to the sigmoid function with $C=500$, as shown in Figure~\ref{fig:sigmoid_cw}. Thus, the sigmoid
function may not be a unique choice to satisfy {\bf \em R1, R2}, and {\bf
\em R3}. However, the sigmoid function seems to be more natural, if we
consider the possible scenario that the 802.11 chipset allows more choices of CW
size than $2^i$ values. (ii) In theory, the access probability similar to
that in \eqref{eq:initial_access} has been used in continuous
\cite{SRS09} as well as discrete time framework \cite{qcsma}, referred
to as {\em Glauber dynamics}. Such a choice of access probability and
transmission length is just one of the combinations. However, the
access probability in \eqref{eq:initial_access} is practically important
for high performance, where the choice of $C$ is crucial.

% \begin{remark}[Glauber dynamics]
%   In theory, the access probability similar to that in
%   \eqref{eq:initial_access} (i.e., $C=1$) has been used in continuous
%   \cite{SRS09} as well as discrete time framework \cite{qcsma}, referred
%   to as Glauber dynamics. Such a choice of access probability and
%   transmission length is just one of the combinations, where other combinations
%   can work in their framework. However, in this paper, we claim that the
%   access probability in \eqref{eq:initial_access} is practically important
%   for high performance.
%   % Central to oCSMA algorithms is the so-called \emph{Glauber dynamics},
%   % which is a Markov chain that can be used to sample the independent
%   % sets of a graph according to a product-form distribution. The
%   % algorithm used in \cite{mixing11, qcsma} is a generalization of
%   % Glauber dynamics where multiple links are allowed to update their
%   % states in parallel. Also the CSMA algorithms used in \cite{LJ08,
%   %   SRS09} can be viewed as continuous-time Glauber dynamics with
%   % adaptive parameters.  More specifically, the Glauber dynamics
%   % algorithm takes the access probability of
%   % $\frac{\exp(q_l)}{\exp(q_l)+1}$ and the holding time of $\exp(q_l)+1$
%   % for all links $l$, which can be the theoretical motivation of our
%   % approach.
% \end{remark}

\subsubsection{Mixture of Contention Levels}
\label{subsubsec:fim_fc}
%\medskip
%\noindent{\bf \em Topological conflicts: Both FIM and FC}
%\smallskip

\begin{figure}[t!]
  \centering
  \subfigure[FIM with FC]{\includegraphics[width=0.48\columnwidth]{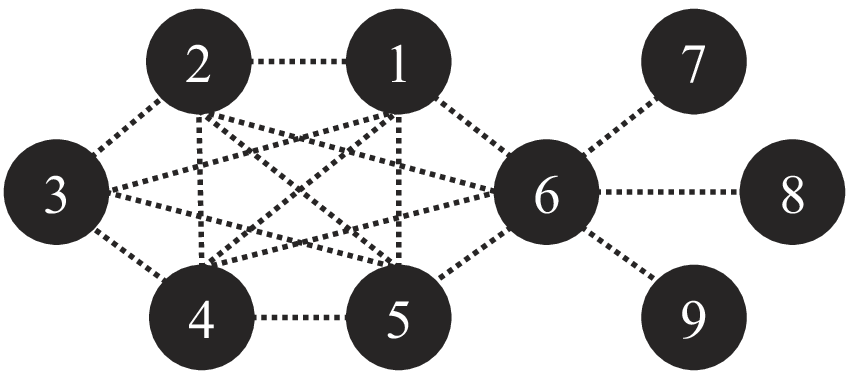}
    \label{fig:star_with_fc}} \hspace{0.2in}
  \subfigure[FC in FIM]{\includegraphics[width=0.23\columnwidth]{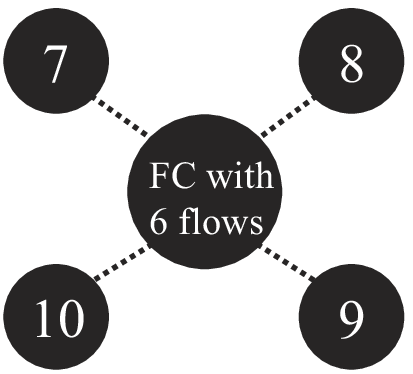}
    \label{fig:fc_in_star}}
  \caption{Mixed topologies' conflicting graphs; (a) The flow 6 belongs to both FC and FIM topologies;
    % , while the flow indexed by 7 belongs to star, both of which interfere with each other;
    (b) six flows within a FC group form a FIM topology with four outer flows.}
  \label{fig:mixed_topologies}
\end{figure}

In practice, it is possible for a link to appear in a mixture of topologies 
with symmetric and asymmetric contention. % At this time, it may seem that BEB and sigmoidal initial
% CW selection conflict with each other.
We study this issue using the example scenarios in
Figure~\ref{fig:mixed_topologies}.  First, in
Figure~\ref{fig:star_with_fc}, flow 6 interferes with flows 1, \ldots, 5
in a fully-connected fashion, and also with flows 7, 8, and 9 in a
FIM-like fashion.  Since it senses the transmission of both the
remaining links forming the FC group and outer links forming the FIM
group, its queue temporarily builds up, thus having a larger access
probability than outer links due to the sigmoidal curve. However, it
shares the medium equally with others in FC group so that their queue
lengths increase together. Thus, in the worst case, BEB can prevent too
aggressive access among the links within FC group. More importantly,
{\em even the reduced access probability from BEB is kept larger than
  those of outer links} (see Section~\ref{subsec:contention} for
simulation results), thus still being sufficiently prioritized in the
FIM topology, preserving proportional fairness. Similar trends are also
observed in Figure~\ref{fig:fc_in_star}.

%%% Local Variables:
%%% mode: latex
%%% TeX-master: "main"
%%% End:

%\subsection{More Features of O-DCF}
\subsection{Imperfect Sensing and Capture Effect}\label{subsec:sensing}

\begin{figure}[t!]
  \centering
  \subfigure[HT]{\includegraphics[width=0.22\columnwidth]{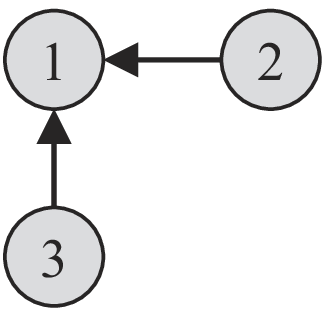}
    \label{fig:ht}} \hspace{0.2in}
  \subfigure[IA]{\includegraphics[width=0.22\columnwidth]{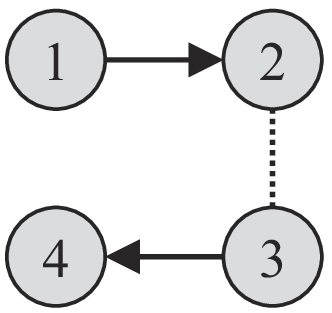}
    \label{fig:ia}} \hspace{0.2in}
  \subfigure[HT w/ cap.]{\includegraphics[width=0.22\columnwidth]{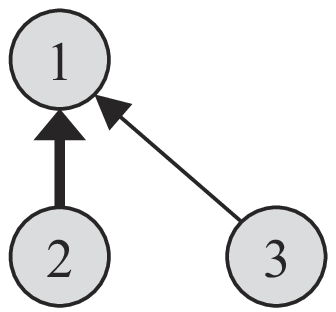}
    \label{fig:ht_cap}} \hspace{0.1in}
%  \subfigure[FC w/ cap.]{\includegraphics[width=0.2\columnwidth]{figure/fc_capture.eps}
%    \label{fig:fc_cap}}
  \caption{Example topologies of imperfect sensing and packet capture.
  In network graphs, vertices represent nodes, dotted lines represent connectivity, and 
  arrows represent flows. (a) HT; (b) IA; (c) HT with capture: a thin arrow for weak signal.}
  %; (d) FC with capture.}
  \label{fig:ht_ia_topo}
\end{figure}

% \noindent{\bf \em Dynamics under excessive collisions.}
We have so far explained why O-DCF works with a focus on the case when
sensing is perfect. However, in practice, sensing is often {\em imperfect},
whose cases are discussed under the scenarios, illustrated
in Figure~\ref{fig:ht_ia_topo}. Recall that we propose to use the RTS/CTS
based virtual sensing in O-DCF.  However, the impact of imperfect
sensing can still be serious in oCSMA, because of the possible
aggravation cycle that collisions increase queue lengths, which in
turn leads to more aggressive access and thus heavier collisions,
especially under the CW adaptation.
%if the transmission intensity is controlled only by $p_l$ for each link $l$.

Our queue based initial CW with BEB substantially lessens such bad
impacts. In {\em HT}, if the queue lengths of
hidden nodes are large, BEB lets each node increase its CW, so that with
small time cost, a transmission succeeds. This successful transmission
generally decreases the queue lengths, preventing CW from being too
small due to queue based initial CW selection.  In {\em IA}, collisions
are asymmetric. Suppose that at some time the advantaged and the
disadvantaged flows have small and large queue lengths,
respectively. Then, the disadvantaged flow will have a smaller initial
CW, thus leading to some successful transmissions and simultaneously
the advantaged flows will hear CTS signaling (responding to RTS from the
disadvantaged flow) and stop their attempts. This helps a lot in
providing fairness between two asymmetric flows.  The packet capture
effects can be handled by our method similarly to the IA scenario, i.e.,
interference asymmetry.  For example, in {\em HT with capture,} similar
behaviors to the IA case occur between weak and strong nodes.
%The capture may
%happen over even in the FC scenario, {\em FC with capture}.
%The weak node builds up its queue, compared to the strong node, which makes
%them have different initial CWs. Hence, the weak node tries to access more
%aggressively so that it instantly receives more service than the strong node,
%and then restores its throughput.

%%% Local Variables:
%%% mode: latex
%%% TeX-master: "main"
%%% End:

\section{O-DCF Implementation}\label{sec:implement}
%\subsection{Other Issues for Compatibility with 802.11 Chipset} \label{sec:implement}

\subsection{Queue Structure}
We implemented O-DCF through an overlay MAC over legacy 802.11
hardware using our C-based software platform, which
requires protocol implementation on top of MadWiFi device driver
\cite{madwifi}.
%consists of GloMoSim \cite{glomosim} based simulator and
%protocol implementation on top of MadWiFi device driver \cite{madwifi}.
Due to a limited memory size of legacy network interface card (NIC), we implement CQ and MAQ at the
user space level.  Note that the scheduling from MAQ to IQ is not an
actual packet transmission to the media ({\bf \em C\ref{const1}}). To
minimize the temporal gap between the service from MAQ and the actual
transmission, we reduce the buffer limit of IQ to one through a device
driver modification.
%Thus, we only require some device driver modification to adapt CSMA parameters
%on every packet as requested by O-DCF scheduler.
%We proceed to describe the main components of our MadWiFi modification.

\subsection{O-DCF Scheduler and Parameter Control}
A scheduler in O-DCF schedules the packets enqueued at MAQ to send them
into the 802.11 hardware, as shown in
Figure~\ref{fig:queue_structure}. Whenever IQ in 802.11 becomes
empty, the scheduler in O-DCF is notified by a system call such as {\tt
  raw socket} function and determines the next packet to be enqueued into
IQ, by comparing the sizes of multiple per-neighbor MAQs. Meanwhile, the
scheduler maintains CSMA parameters, such as CW, AIFS, and NAV values, for
each MAQ's HOL packet. To facilitate packet-by-packet parameter control,
we piggyback such parameters into the header of HOL packet, so that the
modified driver can interpret and set them in the TXQ descriptor of
an outgoing packet for the actual transmission.

%\subsubsection{Compatibility with 802.11 Hardware} \label{subsubsec:hardware}
%\noindent {\bf \em Long data transmission.}
\subsection{Long Data Transmission}
When packet aggregation is not supported in legacy 802.11 hardware
such as 802.11a/b/g, we take the following approach for consecutive
multi-packet transmissions:
The O-DCF scheduler assigns different arbitration inter-frame spaces
(AIFSs) and CWs for the packets inside the specified transmission
length. Since AIFS defines a default interval between packet
transmissions and the smallest CW indicates the shortest backoff time,
this provides a prioritization for {\em back-to-back} transmissions
until the given transmission length expires.  Further, we exploit the
network allocation vector (NAV) option that includes the time during
which neighbors remain silent irrespective of sensing. This
guarantees that even interfering neighbors that cannot sense (due to,
e.g., channel fluctuations) do not prevent the transmission during the
reserved transmission length. In this way we overcome the constraint on
the maximum aggregate frame size ({\bf \em C\ref{const4}}).
RTS/CTS signaling is conducted
only for the first packet within the given transmission length. By modifying the
device driver, we can turn on or off such a signaling according to the
number of transmitted packets specified by the transmission length.

%Also, unlike
%in the standard 802.11, RTS/CTS signaling should be conducted only for
%the first packet inside the holding time. We modify the device driver so
%that RTS/CTS signaling can be turned on or off according to the number
%of transmitted packets guided by the holding time.

%\smallskip
%\noindent {\bf \em Link capacity update.}
\subsection{Link Capacity Update} \label{subsec:link_update}
%Rate adaptation allows for each device to adapt the runtime transmission rate based
%on the channel condition. It exploits the PHY multi-rate capability and enables
%each device to select the best rate out of the mandated options. To maximize the
%transmission throughput perceived by the receiver, 802.11 devices often turn on their rate
%adaptation (by default) to adjust the runtime transmission rate soley based on the ACK,
%which is sent upon successful delivery of a data packet.
As discussed in Section \ref{subsubsec:channel}, to exploit
multi-rate capability, we need to get the runtime link capacity information from the
device driver.  To fetch the runtime capacity, we periodically examine \texttt{/proc}
filesystem in Linux. The period has some tradeoff between accuracy and overhead.
We employ the exponential moving average each second to smooth out
the channel variations as well as to avoid too much overhead of reading
\texttt{/proc} interface.

\section{Performance Evaluation} \label{sec:experiment}

\begin{figure}[t!]
\centering
    \subfigure[Mobile testbed node]{
        \includegraphics*[width=0.39\columnwidth]{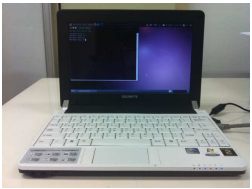}
        \label{fig:Experimental_testbed_KAIST1}
    } \hspace{0.2in}
    \subfigure[FC deployment]{
        \includegraphics*[width=0.43\columnwidth]{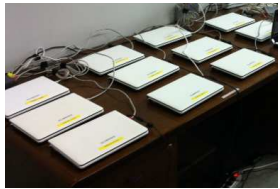}
        \label{fig:Experimental_testbed_KAIST2}
    }
    \caption{Testbed hardware and an example deployment.}
    \label{fig:testbed}
\end{figure}

\subsection{Hardware Testbed}

\noindent{\bf \em Setup.} To evaluate O-DCF's performance, we use both GloMoSim \cite{glomosim}
based simulation and real experiment on a 16-node wireless multihop testbed.
%In this section, we use simulation and experiment based on a 15-node wireless testbed.
%In simulation, we use GloMoSim \cite{glomosim}. In the testbed, we used 15 typical netbook
%platforms (1.66 GHz CPU and 1 GB RAM), as shown in Figure~\ref{fig:testbed}.
In the testbed, each node runs on Linux kernel 2.6.31 and is a netbook
platform (1.66 GHz CPU and 1 GB RAM) equipped with a single 802.11a/b/g
NIC (Atheros AR5006 chipset), as shown in
Figure~\ref{fig:testbed}. We built our O-DCF on top of legacy 802.11
hardware, and modified the MadWiFi driver for O-DCF's operations.
%\note{Relation between repeatability and retry limit or 5 GHz?}
To avoid external interference, we select a 5.805 GHz band in 802.11a.
Also we set the MAC retry limit to four. The default link capacity is
fixed with 6 Mb/s, but we vary the capacity or turn on auto rate adjustment
for the evaluation of the heterogeneous channel cases. For all tested
scenarios, we repeat ten times (each lasting for 100 seconds) and measure
the goodput received at the flow's destination. The length of error bars
in all plots represents standard deviation.
% Other parameters are presented in Table~\ref{table:setup}.
The packet size is set to be 1000 bytes. We choose the $b=0.01$ in
Section~\ref{subsubsec:queue_structure}, and the lower and upper bounds
for MAQ as $Q_{\text{min}}=1$ and $Q_{\text{max}}=1000,$ which, however,
we observe, does not significantly impact the results and the trends in
all of the results.

% \begin{table}[h!]
% \caption{Experimental setup}
% \label{table:setup}
% \vspace{-0.5cm}
%  \begin{center}
%    \begin{tabular}{|c|c|} \hline
%         {\it PHY} & 802.11a, 5.805 GHz band\\ \hline
%         {\it Packet size} & 1000 bytes \\ \hline
%         {\it Traffic pattern} & Concurrent, fully-backlogged flows \\ \hline
%         {\it Utility function} & $\log(x)$ \\ \hline
%         {\it Step size} & $b=0.01$ \\ \hline
%         {\it Virtual queue bound} & $q_{\min}=0$, $q_{\max}=10$ \\ \hline
%  \end{tabular}
%  \end{center}
% \end{table}

\smallskip
\noindent{\bf \em Tested protocols.} We compare the following algorithms:
(i) 802.11 DCF, (ii) standard oCSMA, and (iii) DiffQ \cite{ASSI09}. For
the standard oCSMA algorithm, we implement two practical versions: (i)
{\em CW adaptation} in which we typically fix the transmission length
$\mu$ with a single packet (in our case, 1000 bytes is used and corresponds
to about 150 mini-slots in 802.11a at 6 Mb/s rate) and control the access
probability $p_l(t)$, such that $p_l(t) \times \mu = \exp(q_l(t))$, as used
in the oCSMA evaluation research \cite{bjkysem11}, and (ii) {\em $\mu$ adaptation
with BEB} (simply, $\mu$ adaptation) in which we delegate the selection of
$p_l(t)$ to 802.11 DCF and control $\mu_l(t) = \exp(q_l(t))/p$. Thus, CW
starts from $CW_{\min}=15$ slots in 802.11a and doubles per each collision.
%Note that too large CW values will lead to impractically long transmission lengths under $\mu$ adaptation.
DiffQ is a {\em heuristic} queue-based algorithm using 802.11e feature
and schedules the interfering links with different priorities based on queue
lengths.
Although it shares the philosophy of fair channel access by granting more
chances to less served links with oCSMA and O-DCF, it adapts using several heuristics,
thus guaranteeing suboptimal performance, which makes a difference from oCSMA (under ideal
condition) and O-DCF (even in practice).
%We emphasize that O-DCF is a fundamentally different approach with DiffQ in the sense
%that contention resolution is derived strictly based on
To be consistent, all the protocols evaluated in our testbed are implemented
under 802.11 constraint and thus they keep intrinsic features such as
BEB\footnote{In literature, e.g., \cite{freemac}, BEB is known to be disabled
by using TXQ descriptor in the device driver, which however does not work
in our chipset. We confirmed this via kernel level measurement.}.
%We comment that vanilla oCSMAs in the simulation differ from those in
%the real implementation in our testbed, since there exist some intrinsic
%features in the legacy 802.11 hardware that cannot be disabled, e.g., BEB ({\bf C2}).

\smallskip
\noindent{\bf \em Metrics.} For simple topologies, we directly compute
the optimal rate allocations in terms of proportional fairness and
compare it with the results of the tested algorithms.  In more
large-scale scenarios, it is hard to benchmark the optimal results.
Thus, we use the following method: for $n$ flows and average per-flow
throughput $\gamma_i, 1\leq i \leq n$, we define (i) sum of $\log$
utility for efficiency, i.e., $\Sigma_{i=1}^{n}\log(\gamma_i)$,
and (ii) Jain's index for fairness, i.e., $(\Sigma_{i=1}^{n}
\gamma_i)^2/(n\Sigma_{i=1}^{n} \gamma_i^2)$ \cite{JAIN84}.
\begin{table}[t!]
\caption{Simulation (up) and experiment (down). Aggregate throughput
    in FC topology for 3, 6, 9, 12 flows}
\label{table:fc_result}
\fontsize{8}{10} \selectfont
\vspace{-0.3cm}
 \begin{center}
\begin{tabular}{|c||c|c|c|c|c|}
\hline
$N$&802.11&DiffQ&CW adapt.&$\mu$ adapt.&O-DCF\\
\hline\hline
$3$&4235&4311&4423&4259&4700\\
$6$&3849&4042&1321&4566&4859\\
$9$&3616&2826&1151&4765&4530\\
$12$&3513&1189&116&4814&4501\\
\hline
\end{tabular}
\end{center}
\vspace{-0.3cm}
\begin{center}
\begin{tabular}{|c||c|c|c|c|c|}
\hline
$N$&802.11&DiffQ&CW adapt.&$\mu$ adapt.&O-DCF\\
\hline\hline
$3$&4472&4719&4541&4469&4843\\
$6$&4119&4585&2186&4457&4664\\
$9$&4126&4296&2362&4789&4680\\
$12$&4134&4055&2373&4849&4794\\
\hline
\end{tabular}
\vspace{-0.3cm}
\end{center}
\end{table}

\subsection{Fully-connected, FIM, and Mixture}
\label{subsec:contention}

\begin{figure}[t!]
\centering
    \subfigure[Per-flow throughput]{
        \includegraphics*[width=0.48\columnwidth]{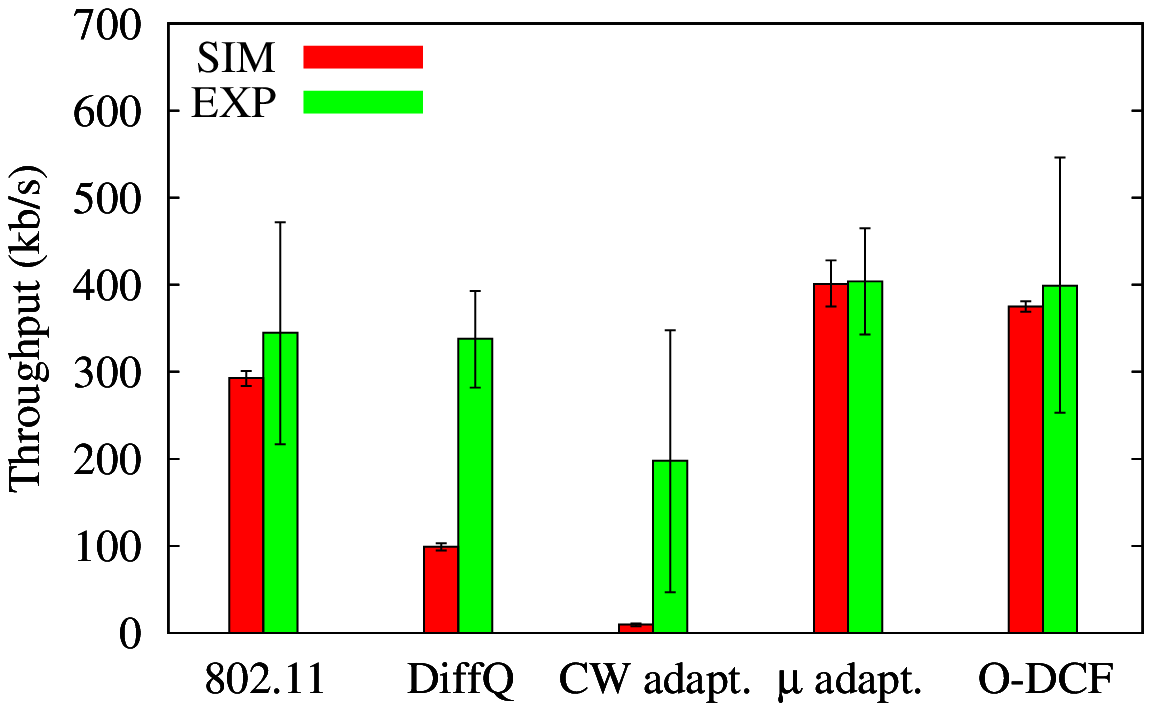}
        \label{fig:fc12f_thruput_comp}
    } \hspace{-0.25cm}
    \subfigure[Collision ratio]{
        \includegraphics*[width=0.48\columnwidth]{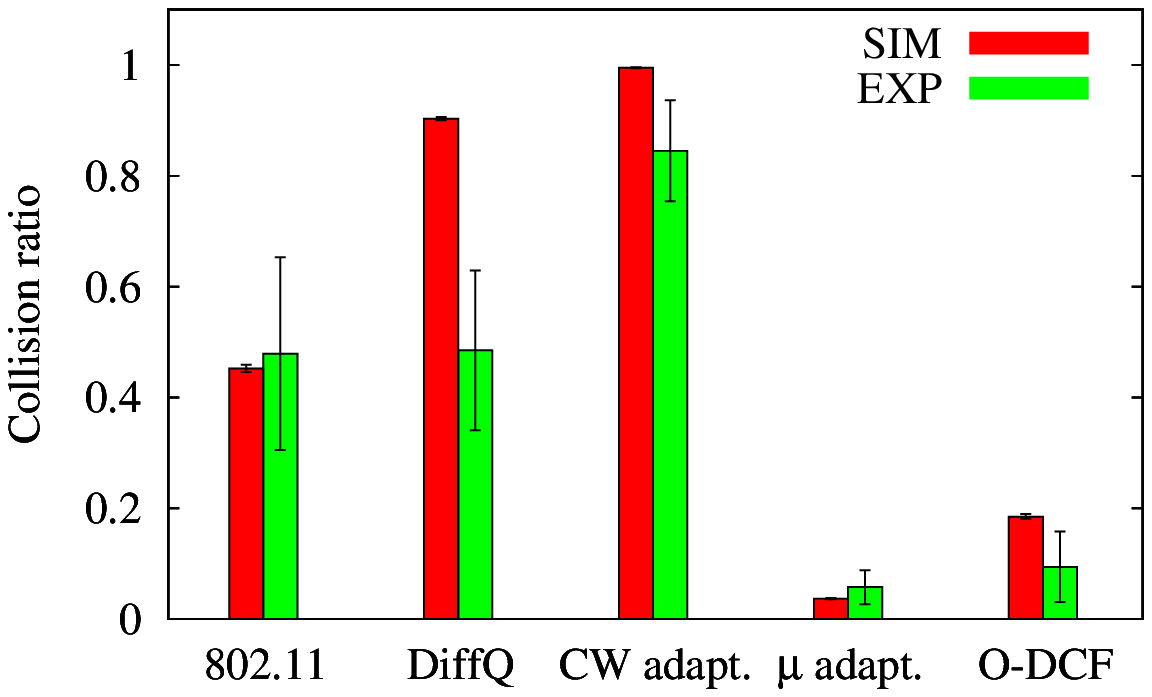}
        \label{fig:fc12f_collision_comp}
    }
    \caption{Performance comparison among tested algorithms in FC topology with 12 flows.}
    \label{fig:fc12f_comp}
\end{figure}

\noindent{\bf \em Fully-connected: Impact of contention degrees.}
We first examine the FC topologies with varying the number of contending
flows (see Figure~\ref{fig:fc4}).  Table~\ref{table:fc_result}
summarizes the aggregate throughputs both in simulation and
experiment. O-DCF outperforms 802.11 DCF, DiffQ, and CW adaptation
regardless of the contention levels. This is mainly due to O-DCF's
adaptivity to the contention levels with BEB and the enlarged
transmission length driven by the queue lengths. CW adaptation and
DiffQ perform poorly, because as the number of contending flows
increases, they have more performance loss, where more collisions lead
to more aggressive access, and thus much more collisions in a vicious cycle.
In 802.11, all nodes start contending with a fixed but aggressive CW value, thus
experiencing quite many collisions despite BEB operation (see
Figure~\ref{fig:fc12f_comp}).

O-DCF shows a little lower throughput than $\mu$
adaptation. It sometimes has a smaller CW than $\mu$ adaptation, causing
more collisions when there are many contending flows. However, such a
difference in CW values makes a big difference of short-term fairness
performance. For example, in FC with 12 flows, we measured the largest
inter-TX time of a flow under two schemes: 3.75s vs. 7.26s for O-DCF and
$\mu$ adaptation, respectively.
%\note{3.25s seems too big for real
%  application. Did we cut by maximum transmission length?}
This can be explained by the corresponding transmission length; O-DCF requires much
smaller (thus practical) transmission length (almost less than 20
packets), while $\mu$ adaptation requires much longer (thus impractical)
transmission length that is often upper bounded by the maximum length
(i.e., 64 packets in our setting).
%Thus, $\mu$ adaptation is seemingly good, but, in fact,
%\note{what does this mean? it does not work as requested by oCSMA.}
We confirm that there is a tradeoff between long-term efficiency and
short-term fairness even in practice.

We comment that DiffQ and CW adaptation in experiment perform better
than that in simulation, because in experiment, BEB is played ({\bf \em C3})
more than its original design (in fact, no BEB operation is allowed in CW adaptation),
which actually shows the power of BEB as a distributed tuning of CW. This is supported
by the fact that 802.11, which has BEB in itself, also performs reasonably well.
%\note{BEB is not included in DiffQ? It uses a 802.11 e, right?}
%However, still, the redesigned oCSMA is better, because we also smartly choose the initial access
%probability based on the queue lengths, so that the initial collisions
%are reduced and time is not wasted much for resolving collisions.
%For example, as shown in Figure \ref{fig:fc12f_comp}, over the FC topology with 12 flows,
%our O-DCF achieves 38.2\% (resp. 22.9\%), 315.7\% (resp. 33\%), and 60.6\% (resp. 7.4\%)
%gain over 802.11 DCF, $p$- and $\mu$-approaches in simulation (resp. experiment), respectively.
The performance difference between experiment and simulation also comes
from the capture effect in experiment in the deployment of nodes in
close proximity as shown in Figure~\ref{fig:Experimental_testbed_KAIST2}.

\begin{figure}[t!]
\centering
    \subfigure[Simulation]{
        \includegraphics*[width=0.48\columnwidth]{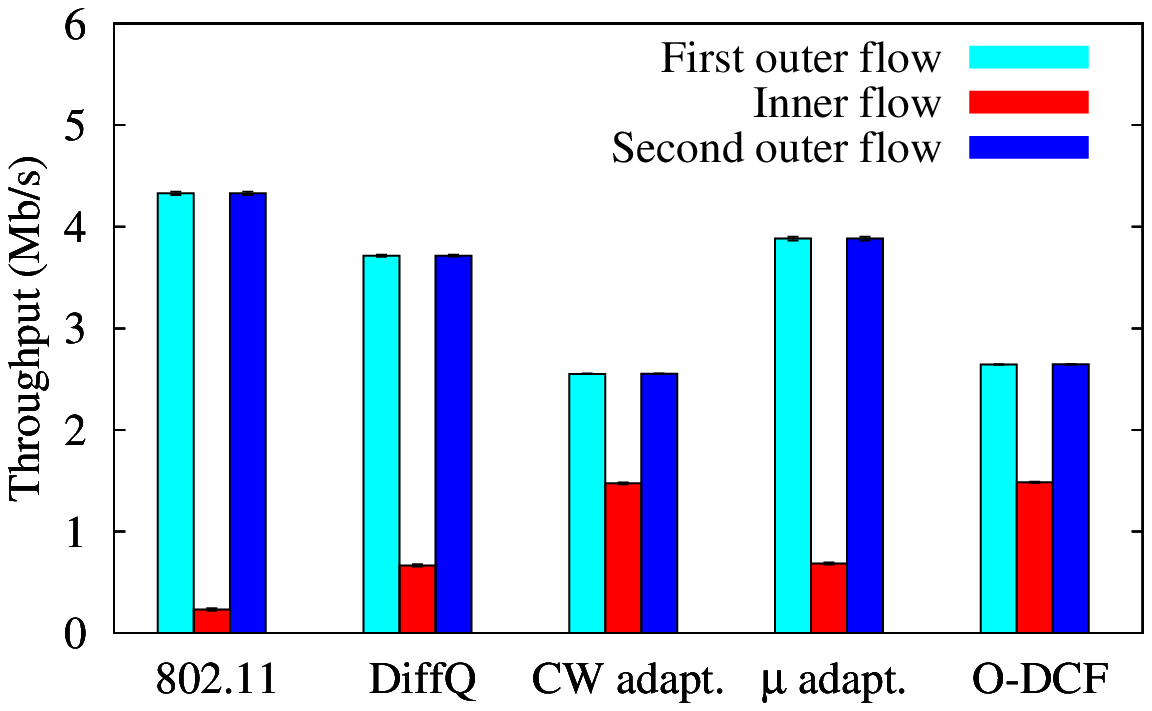}
        \label{fig:sim_fim_thruput_comp}
    } \hspace{-0.25cm}
    \subfigure[Experiment]{
        \includegraphics*[width=0.48\columnwidth]{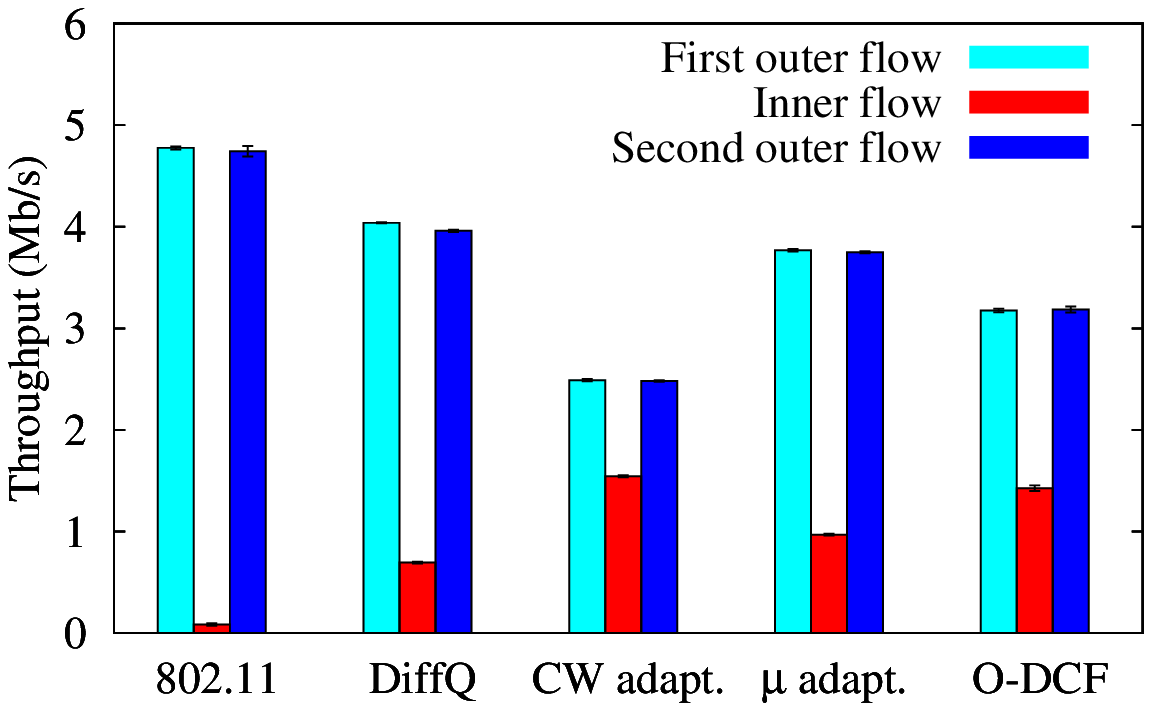}
        \label{fig:exp_fim_collision_comp}
    }
    \caption{Performance comparison among tested algorithms in FIM
      topology with two outer flows.}
    \label{fig:fim_thruput_comp}
%    \vspace{-0.4cm}
\end{figure}

\begin{figure}[t!]
\centering
    \subfigure[3 outer flows]{
        \includegraphics*[width=0.48\columnwidth]{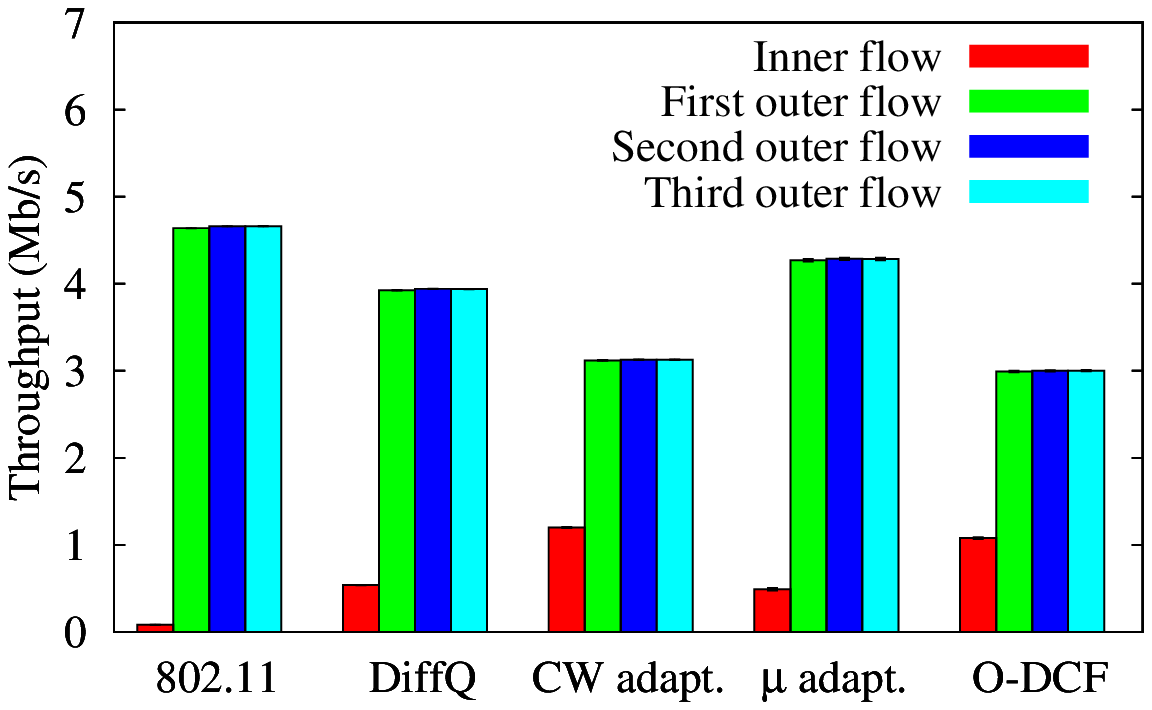}
        \label{fig:sim_star4f_thruput_comp}
    } \hspace{-0.25cm}
    \subfigure[4 outer flows]{
        \includegraphics*[width=0.48\columnwidth]{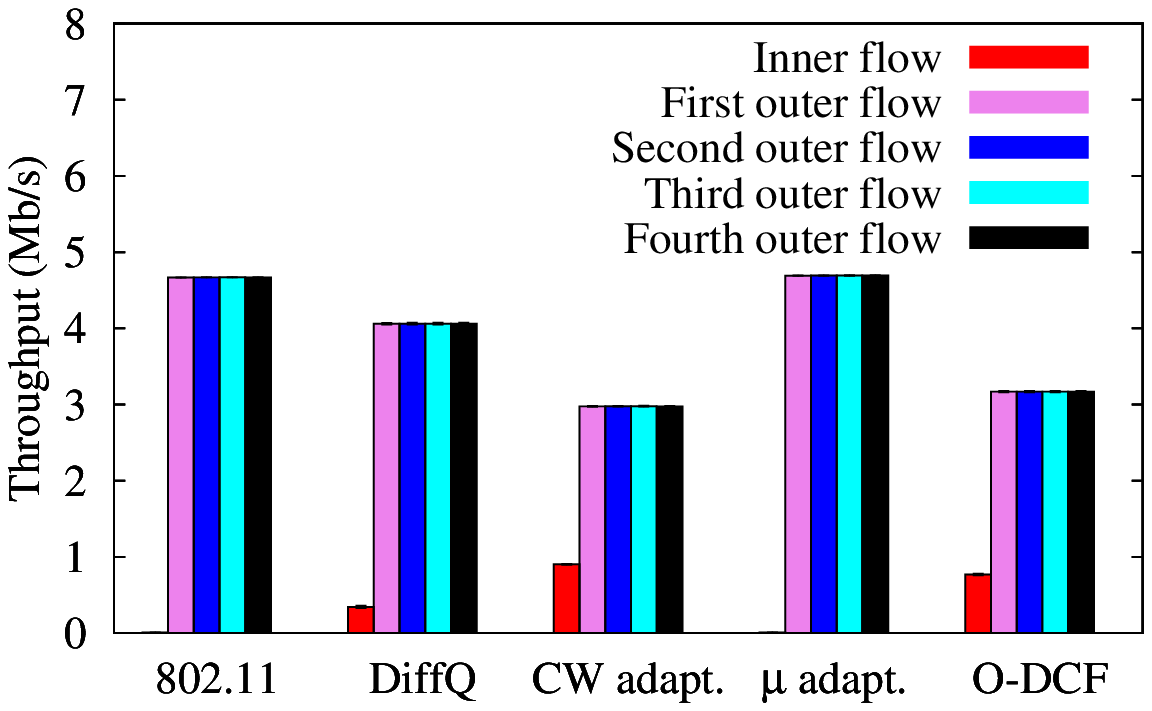}
        \label{fig:sim_star5f_thruput_comp}
    }
    \caption{Simulation. Throughput comparison among tested algorithms in FIM
      topology with three and four outer flows. %\note{why only simulation? difficult to make the topology?}
      }
    \label{fig:sim_star_thruput_comp}
\end{figure}

\smallskip
\noindent{\bf \em FIM: Impact of contention asymmetry.}
Next, we study the case of asymmetric contention, using the FIM
topology (see Figure~\ref{fig:fim}). In this case, it is well-known that
the central flow experiences serious starvation \cite{MTE06, ASSI09,
  WYMS10} in 802.11 DCF, as also shown in
Figure~\ref{fig:fim_thruput_comp}.  As mentioned earlier before, we
observe that collisions are rare (less than 10\%\footnote{This seems
  quite high, but the reason is that the total number of TX attempts at
  central flow is small. Thus, the absolute number of collisions is
  small.} in simulation, less than 2\% in experiment). In contrast
to the FC cases, BEB does not operate frequently in experiment, and thus
the performance results are very similar in simulation and experiment.

A main property to verify is whether an algorithm achieves
proportional-fair (PF) or not.  We can see that O-DCF almost achieves PF exactly,
where 2:1:2 is the optimal throughput ratio under PF in FIM with two outer flows.
This is due to the efficient access differentiation
caused by our queue based initial CW selection. As expected, 802.11's
starvation of the central flow is the most serious.
%Regarding Jain's index, we can improve the fairness of 44.5\% through O-DCF over 802.11 DCF.
DiffQ resolves the starvation of the central flow well, but it shows
suboptimal performance due to the heuristic setting of CW size.  We see
that CW adaptation is also good in fairness, because in presence of few
collisions, CW adaptation also has an adaptive feature of adjusting CW
to the queue lengths. However, in $\mu$ adaptation, there still exists a
lack of fairness between the central and outer flows, because the CW
size is fixed with a small value, similarly to 802.11 and thus the outer
flows have more power to grab the channel. On the other hand, the
central flow needs a longer transmission length once it grabs the
channel. This extended transmission of the inner flow leads to the queue
buildup of outer flows, which, in turn, makes outer flows access the
channel more aggressively. This unfairness is amplified as the
number of outer flows increases, as shown in Figure~\ref{fig:sim_star_thruput_comp},
where we only plot the simulation results since the topologies cannot be easily
produced in a hardware testbed.

\smallskip
\noindent{\bf \em Mixture: Impact of combination of topological features.}
We perform simulations to study how O-DCF works in the topology with
a mixture of FC and FIM, as shown in Figure~\ref{fig:mixed_topologies}.
Earlier, we mentioned that two features (BEB and queue-based initial CW selection) {\em may}
conflict with each other. Now we will show that it actually does not here. % are seemingly
% conflicting.
% mixed networks
% consisting of FC and star where two requirements {\em R2} and {\em R3} are seemingly
% conflicting. As described in Section \ref{subsec:accessprob}, we simulate two distinct
% topologies (Fig. \ref{fig:mixed_topologies}).
% whose proportional fair ratios can be
%numerically calculated (see Table \ref{table:mix_thruput_ratio}).

%\begin{table}[h!]
%\caption{Optimal throughput ratio for two mixed topologies of Fig. \ref{fig:mixed_topologies}.}
%\label{table:mix_thruput_ratio}
%\vspace{-0.7cm}
%\fontsize{8}{10} \selectfont
%\begin{center}
%\begin{tabular}{|c|c|c|c|c|c|c|c|c|c|c|} \hline
%Flow index &1&2&3&4&5&6&7&8&9&10\\ \hline\hline
%Star w/ FC&1.8&1.8&1.8&1.8&1.8&1&9&9&9&N/A\\ \hline
%FC in star&1&1&1&1&1&1&4&4&4&4\\ \hline
%\end{tabular}
%\end{center}
%\vspace{-0.3cm}
%\end{table}

\begin{figure}[t!]
\centering
    \subfigure[FIM with FC (9 flows)]{
        \includegraphics*[width=0.47\columnwidth]{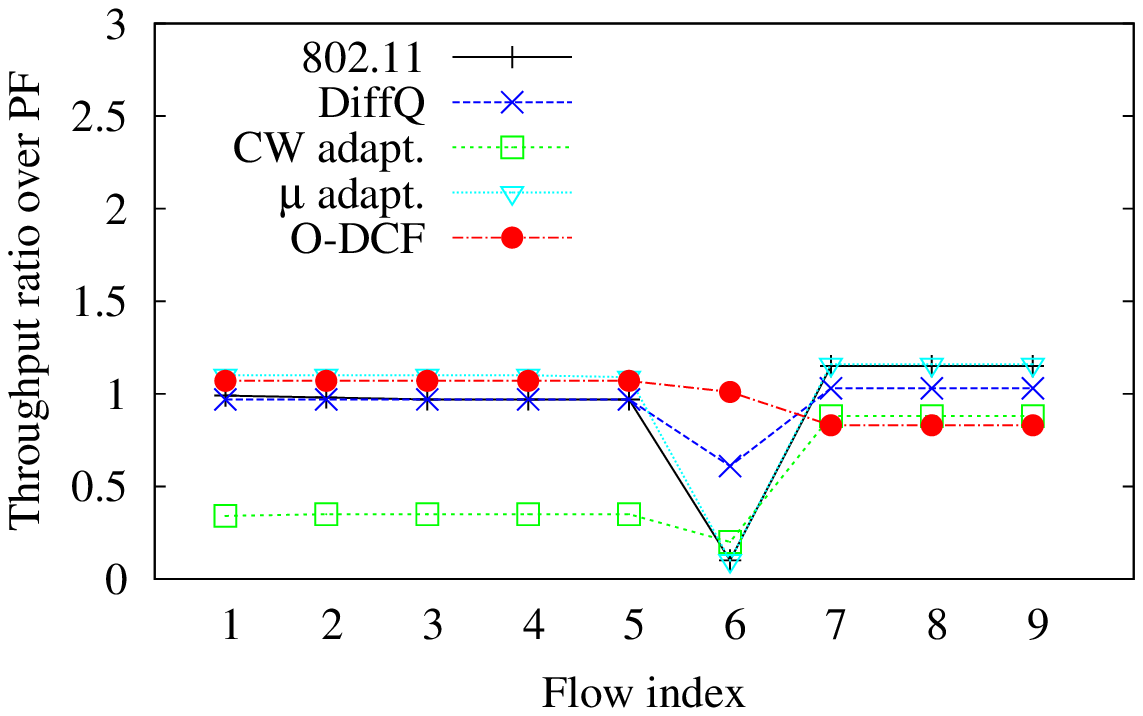}
        \label{fig:sim_mix9f_perflow_thruput}
    }\hspace{-0.1cm}
    \subfigure[FC in FIM (10 flows)]{
        \includegraphics*[width=0.47\columnwidth]{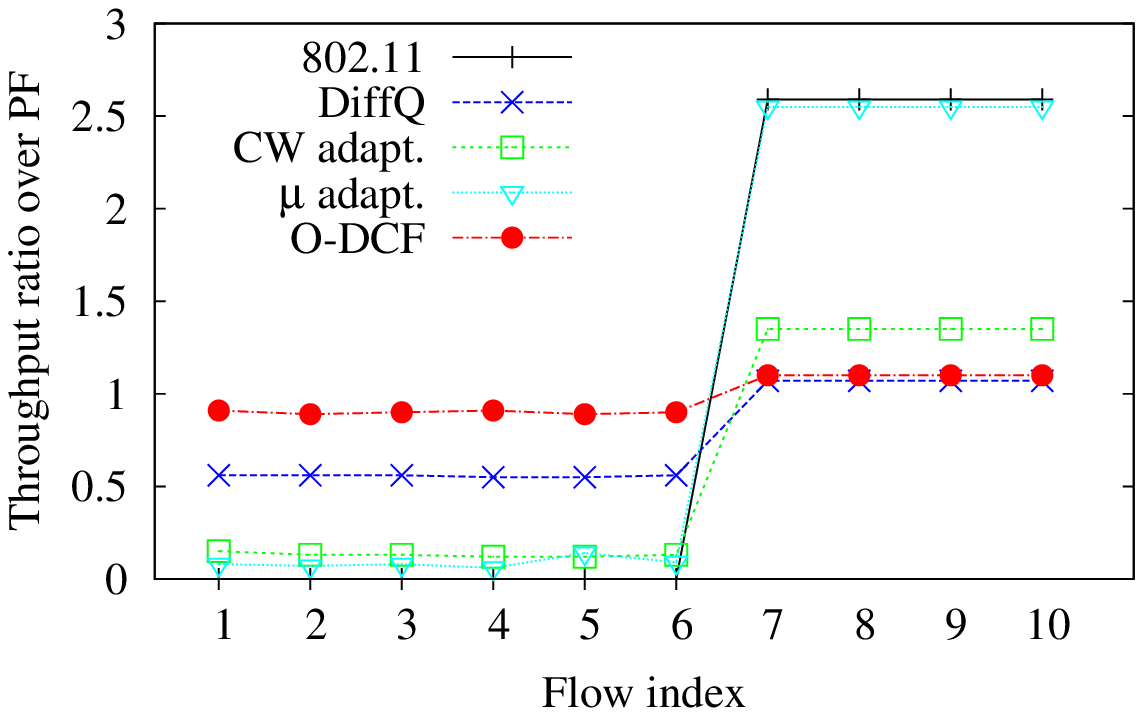}
        \label{fig:sim_mix10f_2_perflow_thruput}
      }
    \subfigure[CWs of O-DCF in ``FIM with FC'' topology]{
        \includegraphics*[width=0.46\columnwidth]{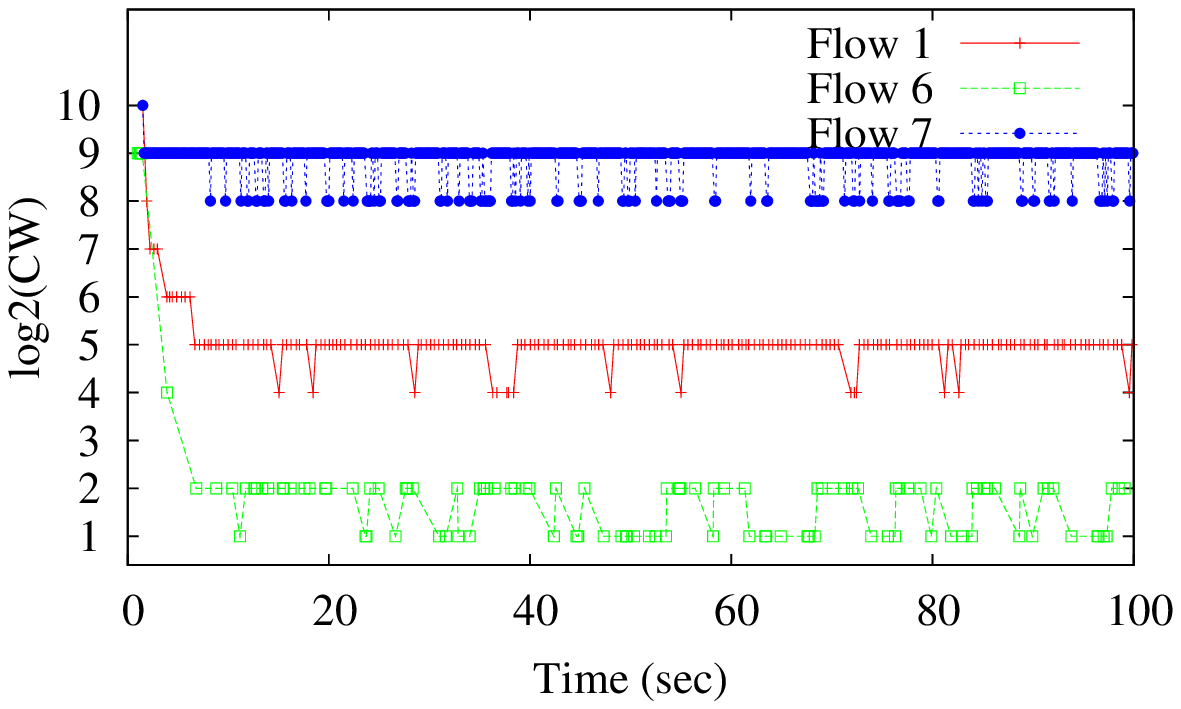}
        \label{fig:sim_mix9f_cw_trace}
    }
    \subfigure[CWs of O-DCF in ``FC in FIM'' topology]{
        \includegraphics*[width=0.46\columnwidth]{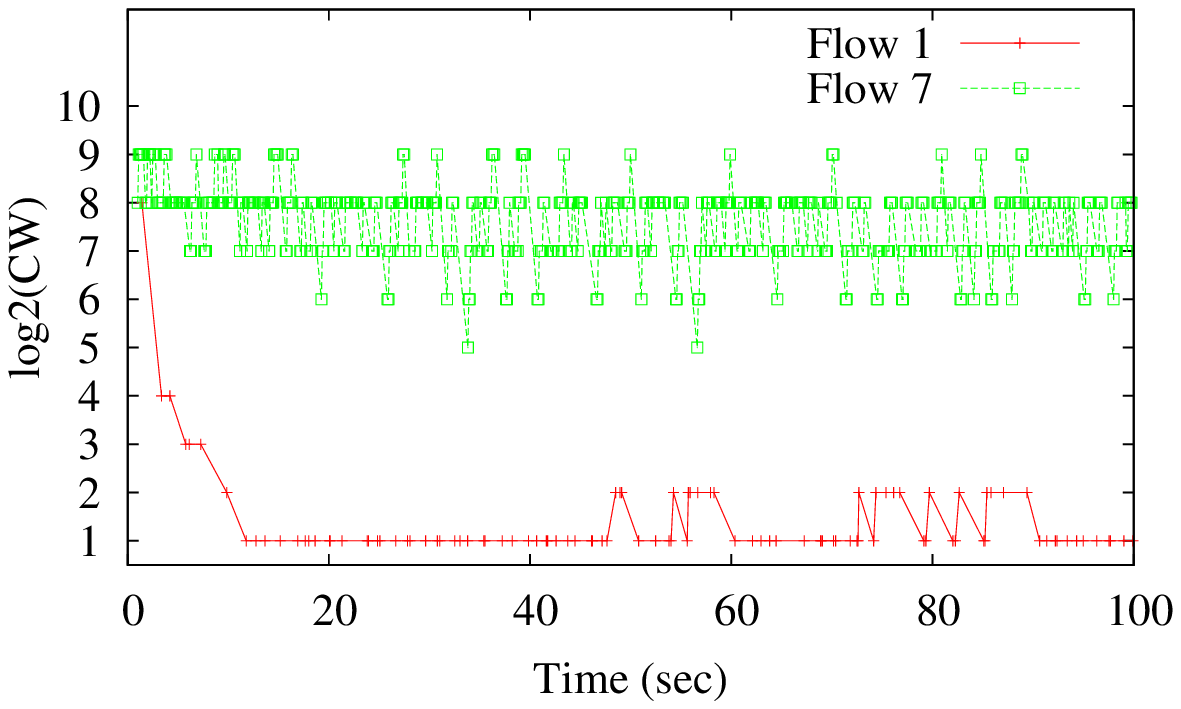}
        \label{fig:sim_mix10f_2_cw_trace}
    }
    \caption{Simulation. (a)-(b): Performance comparison among tested algorithms in two
    mixed topologies; the flow indexes are the same as in Figure~\ref{fig:mixed_topologies};
    we plot the throughput ratio over optimality: the closer to one the ratio is, the better fairness is;
    (c)-(d): CW traces of representative flows are plotted over time in O-DCF. }
    \label{fig:sim_mix_performance}
\end{figure}

Figure~\ref{fig:sim_mix9f_perflow_thruput} shows the normalized per-flow
throughput by the optimal one (i.e., PF share) over the topology in
Figure~\ref{fig:star_with_fc}.  We observe that O-DCF outperforms
others in terms of fairness. The reason is explained as follows: Suppose
that flow 6 is fully-connected with flows 1-5. Then, its access probability
will be set to be some value, say $p_6,$ so that collisions are more or less
minimized. However, in Figure~\ref{fig:star_with_fc}, flow 6 should have a
high access probability to be prioritized over flows 7-9. A desirable solution
would be to make flow 6's probability higher than
$p_6$ (for prioritization considering the FIM part), as well as flows 1-5's
probabilities lower than that in the case of only FC (for collision reduction).
This is indeed achieved by our queue-length based initial CW selection with
BEB in a distributed manner, verified by Figure~\ref{fig:sim_mix9f_cw_trace}.
Clearly, flows 1-5 need longer transmission lengths (and thus hurting short-term fairness).
They turn out not to be impractically long in the scenario considered here.
Similar principles are applied to the topology in Figure~\ref{fig:fc_in_star},
whose CW traces are shown in Figure~\ref{fig:sim_mix10f_2_perflow_thruput}.

\subsection{Imperfect Sensing and Capture Effect}\label{subsec:imperfect_exp}

We now investigate the impact of imperfect sensing in O-DCF with the
topologies: HT, IA, and HT with capture, which are depicted in Figure~\ref{fig:ht_ia_topo}.
As discussed in Section~\ref{subsec:sensing}, we enable a virtual sensing, i.e., RTS/CTS
signaling by default for better channel reservation before data transmission in all tested
algorithms.

\begin{figure}[t!]
\centering
    \subfigure[Simulation]{
        \includegraphics*[width=0.48\columnwidth]{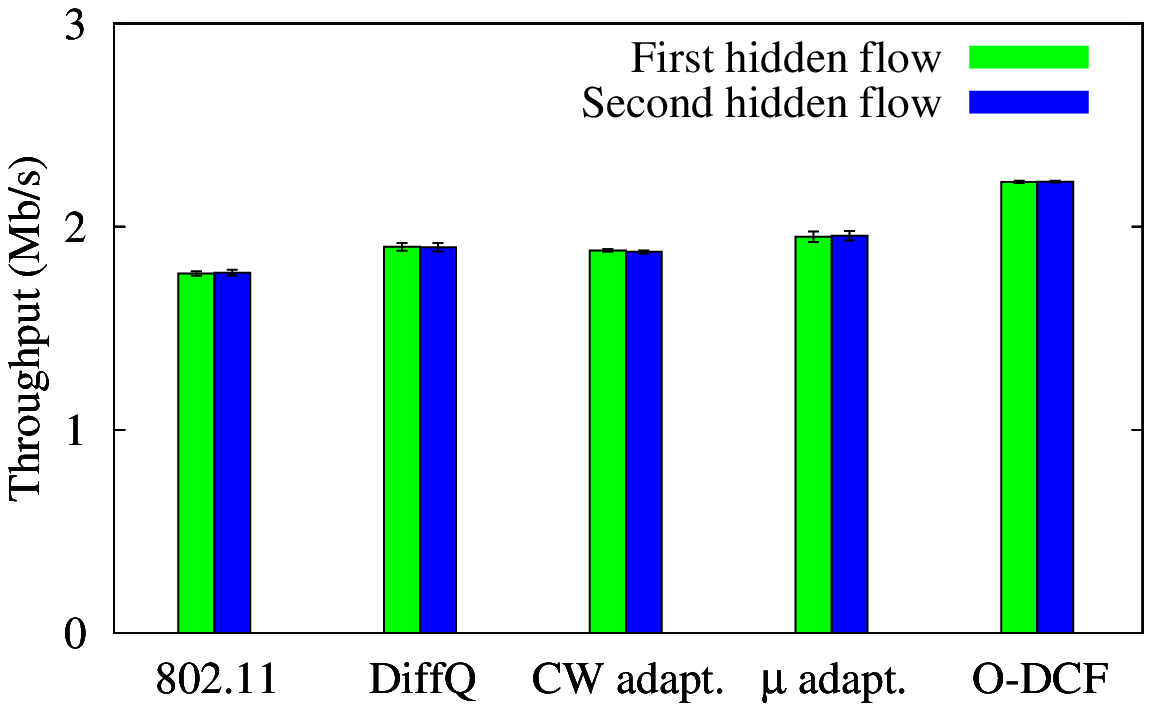}
        \label{fig:sim_ht_thruput_comp}
    }\hspace{-0.2cm}
    \subfigure[Experiment]{
        \includegraphics*[width=0.48\columnwidth]{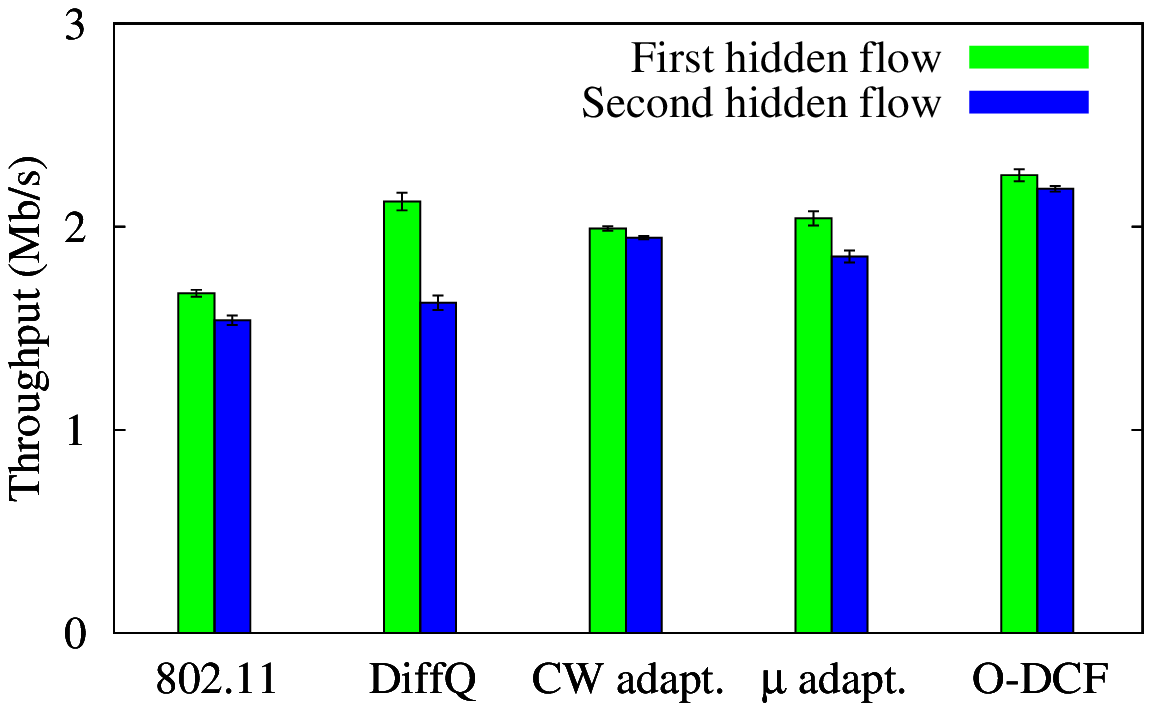}
        \label{fig:exp_ht_collision_comp}
    }
    \subfigure[Experiment in HT w/ cap.]{
        \includegraphics*[width=0.48\columnwidth]{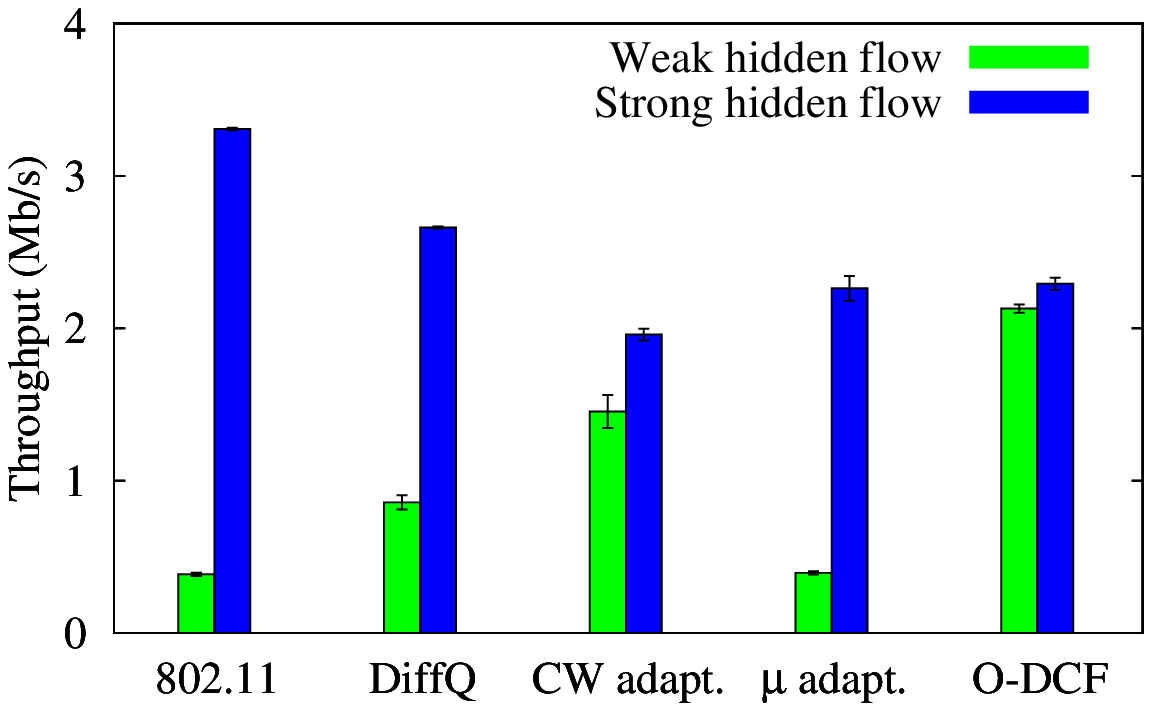}
        \label{fig:exp_ht_asym_collision_comp}
    }
    \caption{Throughput comparison among tested algorithms in HT
      topology w/o (up) and w/ packet capture (down).}
    \label{fig:ht_thruput_comp}
%    \vspace{-0.5cm}
\end{figure}

\begin{figure}[t!]
\centering
    \subfigure[Simulation]{
        \includegraphics*[width=0.48\columnwidth]{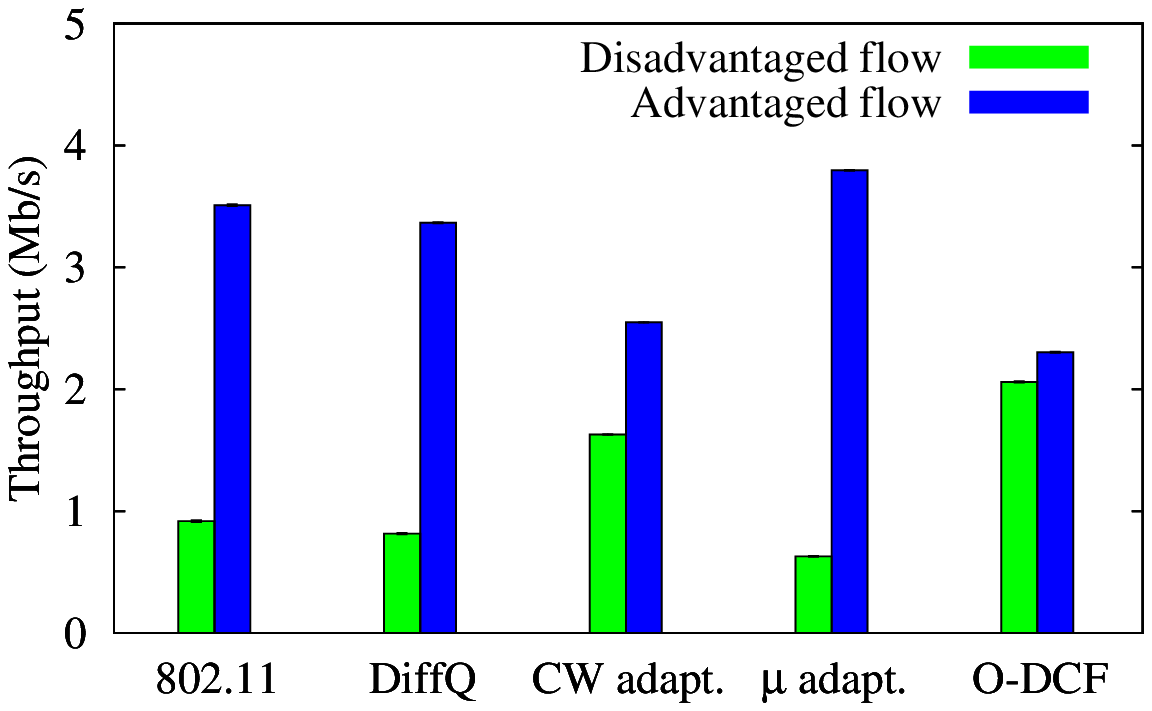}
        \label{fig:sim_ia_thruput_comp}
    }\hspace{-0.2cm}
    \subfigure[Experiment]{
        \includegraphics*[width=0.48\columnwidth]{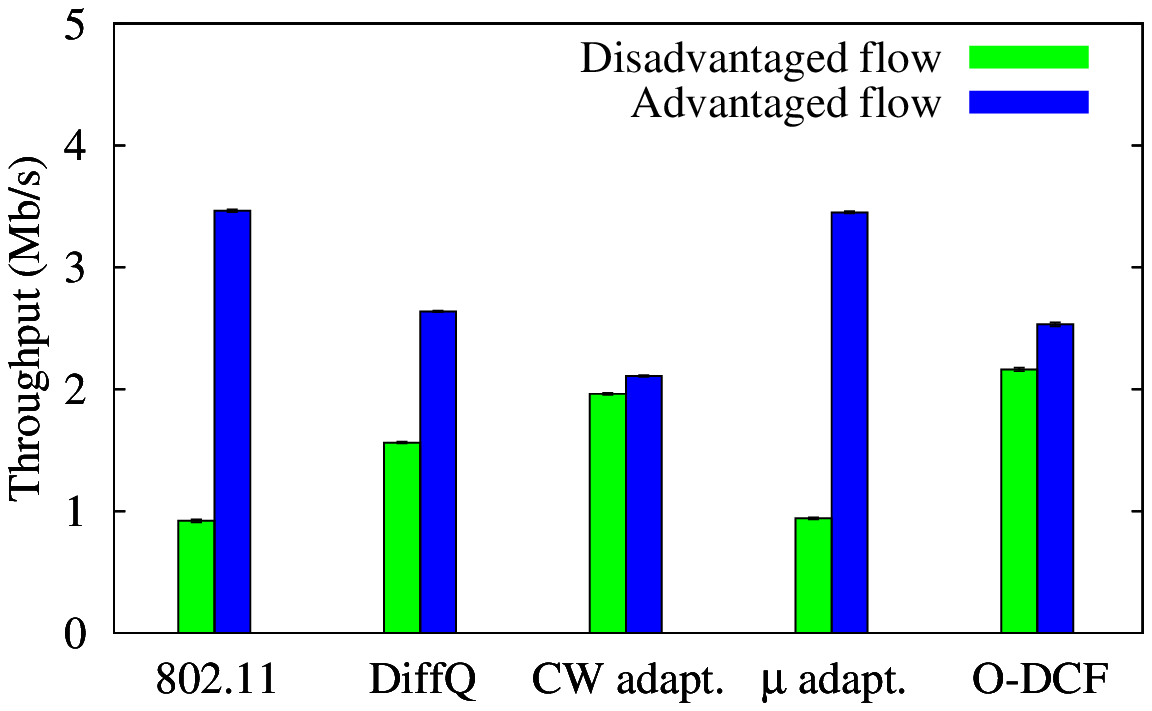}
        \label{fig:exp_ia_collision_comp}
    }
    \caption{Throughput comparison among tested algorithms in IA topology.}
    \label{fig:ia_thruput_comp}
\end{figure}

Figures~\ref{fig:ht_thruput_comp} and \ref{fig:ia_thruput_comp} show the
throughput results for HT and IA, respectively. First, in HT, O-DCF
outperforms others in symmetric interference conditions by hidden nodes,
as well as asymmetric conditions due to packet capture.  Particularly,
in O-DCF, collisions and thus BEB allow the hidden flows to access
the media with larger CWs (i.e., less aggressiveness), resulting in
many successful RTS/CTS exchanges. Fairness is guaranteed by the
transmission length control.
%has much smaller collisions due to larger
%CW values and thus, RTS packets are likely to equally reserve the channel.
Second, in IA, our O-DCF shows very fair and high throughput, where the
advantaged flow has a larger CW due to small backlogs, and the
disadvantaged flow also uses a larger CW due to BEB, responding well to
collisions. As a result, both flows contend with sufficiently large %\note{why similar?}
CW values without heavy collisions, enjoying better
RTS/CTS-based reservation. Similarly, we can ensure fairness in the
scenario with packet capture.
% , leading to few collisions and better
% reservation via RTS/CTS exchange.  Since collisions at the disadvantaged
% flow induce longer transmission length, our method produces a little more gain
% than the advantaged flow (e.g., 9 vs. 4 packets for average holding
% times, respectively).
Note that $\mu$ adaptation performs badly especially under asymmetric
contention such as HT with capture and IA topologies. Similarly to 802.11 DCF, its fixed CW value makes BEB
happen frequently, hindering channel access of the weak hidden node in HT with capture
or the disadvantaged node in IA. DiffQ performs similarly due to the still aggressive setting
of CW values. However, CW adaptation grants more aggressiveness to less-served flows, thus
reducing the throughput gap between the two flows in both scenarios, while still leaving a small 
gap from optimality.

% The
% overall performance trend is similar between simulation and experiment,
% but the default BEB operation in experiment affects the flow's
% performance differently, depending on whether the interference
% relationship is symmetric or asymmetric.
%since IA scenario only arises from the topological factor. The gap is due to the
%default BEB operation in experiment, as explained earlier; since BEB reduces the
%transmission chances of the disadvantaged flow, it drops the throughput more than
%that in simulation.

% the ability of oCSMA to cope with imperfect sensing environments
% using HT and IA scenarios. Fig. \ref{fig:ht_thruput_comp} shows average throughput
% achieved by two hidden nodes in simulation and experiment; the former produces the
% symmetric interference, whereas the latter produces the asymmetric interference in
% our experimental setting due to physical capture over HT topology.

% Fig. \ref{fig:ia_thruput_comp} presents the performance achieved by two flows in IA
% topology, which is another asymmetric interference scenario due to relative positions
% of two flows.

%Note that the significant throughput improvement in the redesigned oCSMA is achieved by
%selecting a good CW value from the sigmoidal mapping between queue size and access probability
%and by exploiting the conventional feature of RTS/CTS signaling whose overhead is not critical
%since transmitted data size is kept quite long under our method.

%%% Local Variables:
%%% mode: latex
%%% TeX-master: "main"
%%% End:

\subsection{Heterogeneous Channels}

\begin{figure}[t!]
  \centering
  \subfigure[Static case]
        {\includegraphics[width=0.24\columnwidth]{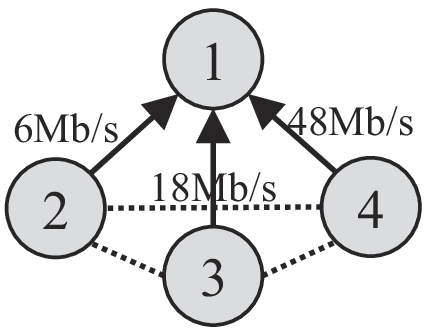}
        \label{fig:hc3}}
  \subfigure[Mobile case]
        {\includegraphics[width=0.24\columnwidth]{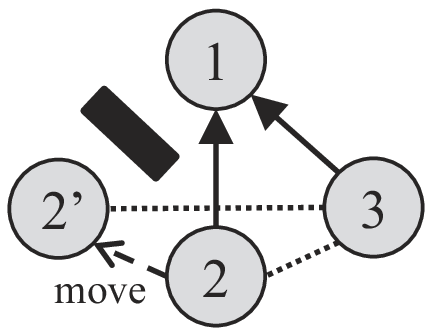}
        \label{fig:hc_move}}
  \subfigure[Per-flow throughput]{
        \includegraphics*[width=0.42\columnwidth]{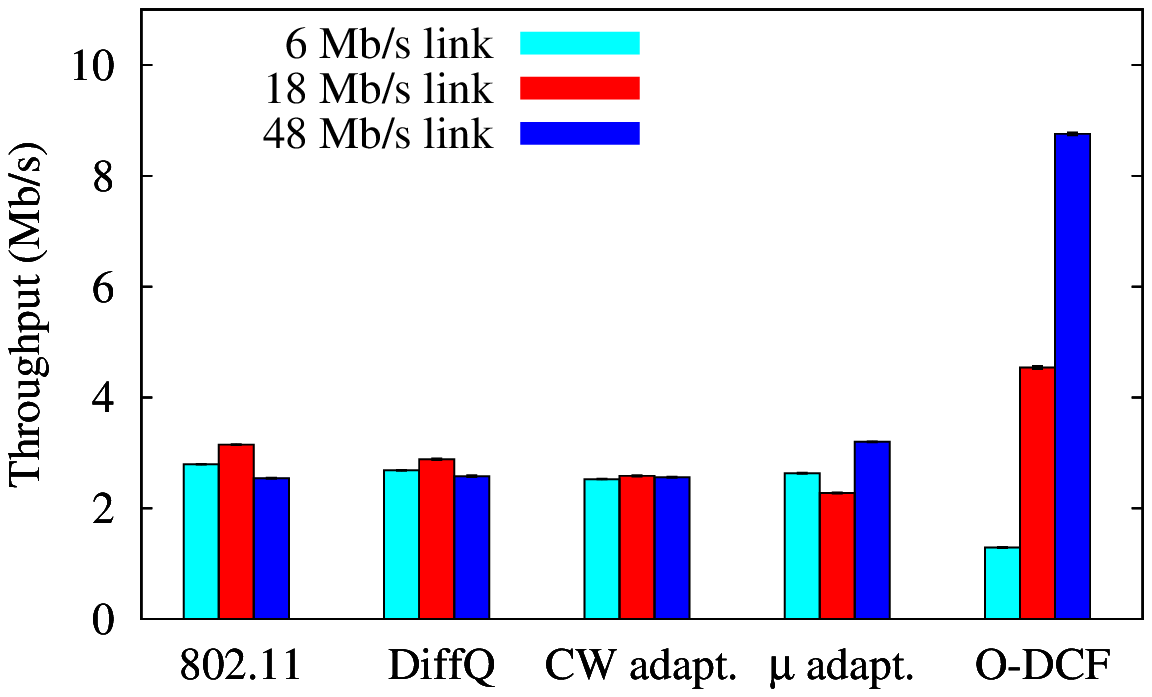}
        \label{fig:exp_hc_thruput_comp}}
  \caption{Experiment. Single-hop network scenario consisting of clients and one AP;
  (a) static case: all nodes remain fixed with different PHY rates; (b) mobile case:
  node 2 moves away from node 1 at 60 seconds.}
  \label{fig:hc_topo}
%  \vspace{-0.5cm}
\end{figure}

We analyze O-DCF's performance for the heterogeneous channel conditions
configured by the topology of Figures~\ref{fig:hc3} and
\ref{fig:hc_move}. We consider two scenarios: (i) {\em static}: nodes
are stationary with different link rates, 6, 18, and 48 Mb/s, and (ii)
{\em mobile}: each node turns on the auto-rate functionality and two
clients send their data to a single AP, and after 60 seconds, one of
two clients (say node 2) moves away from AP (node 1).  In adapting
the rate, we employ the SampleRate algorithm \cite{bicket}, popularly
used in MadWiFi driver. We measure the runtime PHY rate information updated
every one second interval, as mentioned in Section \ref{subsec:link_update}.%\note{Section~\ref{xxx}.}

% We compare the redesigned algorithm described in Algorithm \ref{alg:redesign} with
% (i) vanilla oCSMAs that do not consider the channel information and (ii) 802.11 DCF.
% In the experiment, we consider a single-hop wireless network consisting of multiple
% clients and a single access point (AP), as shown in Fig. \ref{fig:hc3}. For rate
% adaptation, we employ the SampleRate algorithm \cite{bicket}, which is popularly
% used in MadWiFi driver.\footnote{In this paper, we are not concerned about the
% performance of the rate adaptation algorithm itself.} Moreover, we use the runtime
% PHY rate information updated every one second window.

\begin{figure}[t!]
\centering
    \subfigure[Runtime PHY rate]{
        \includegraphics*[width=0.48\columnwidth]{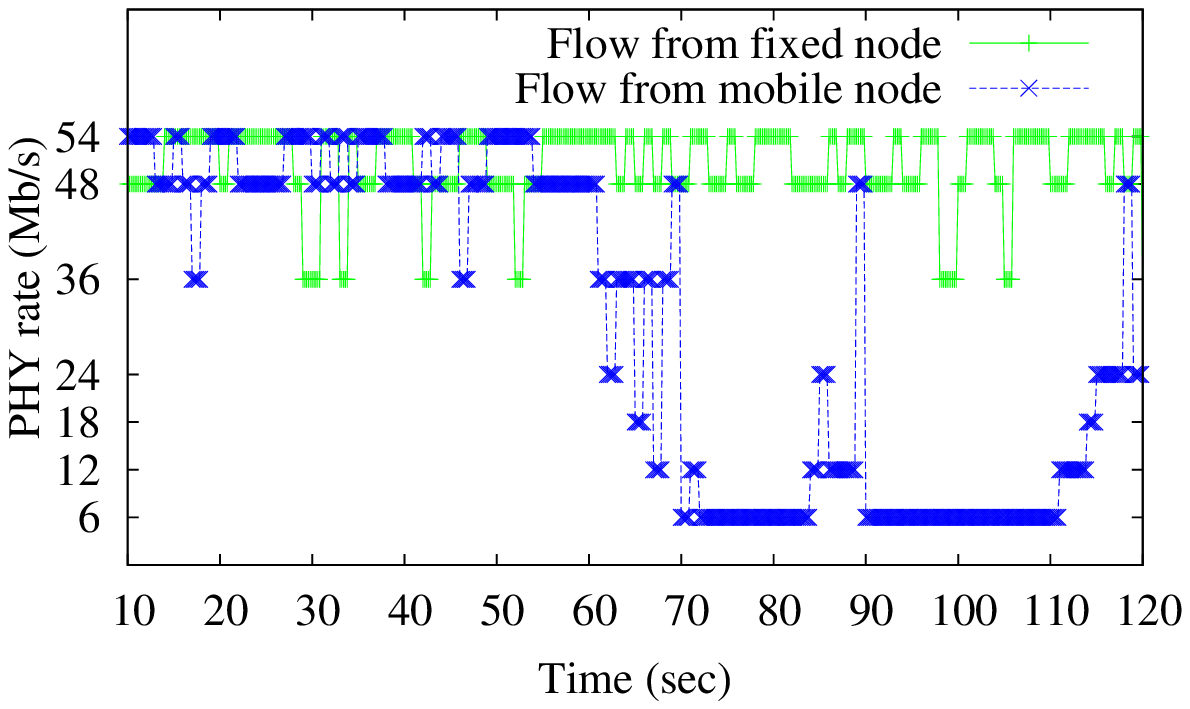}
        \label{fig:exp_hc_auto_rate_trace}
    }\hspace{-0.25cm}
    \subfigure[O-DCF w/ rate update]{
        \includegraphics*[width=0.48\columnwidth]{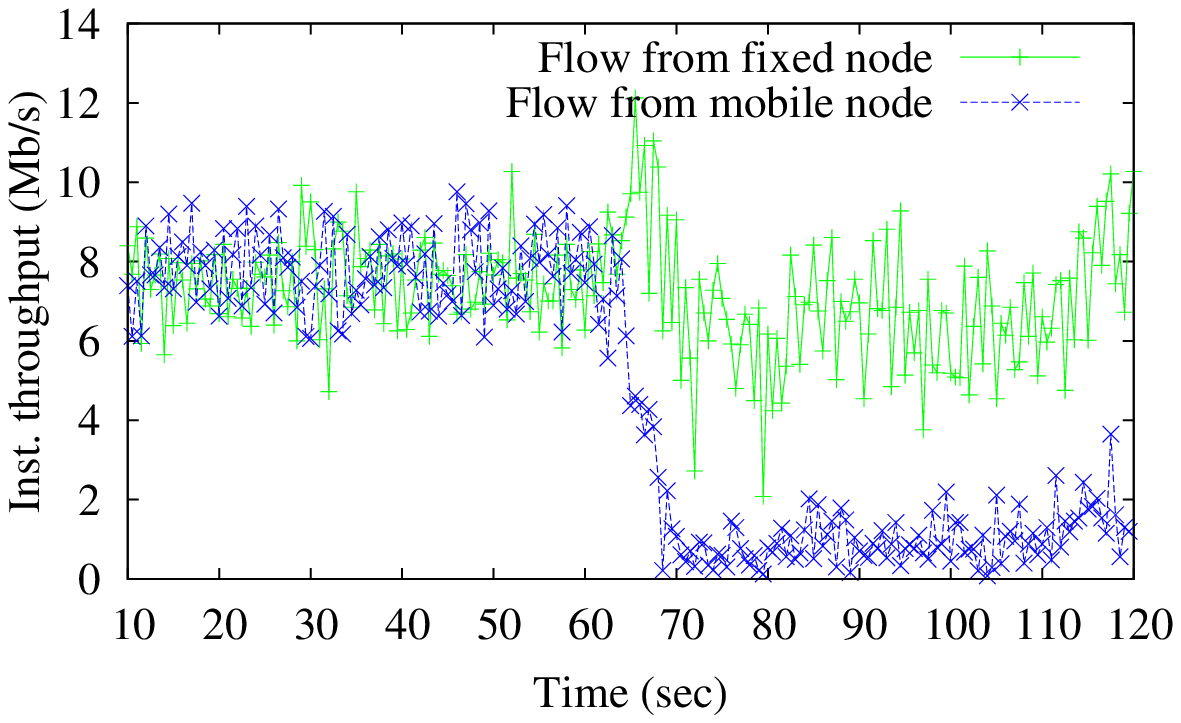}
        \label{fig:inst_thruput_trace_smet}
    }
    \caption{Experiment. Performance evaluation in mobile scenario (Figure~\ref{fig:hc_move}).}
    %single-hop network with two flows in case of node mobility;
    %one client remains fixed while the other client moves away from the AP at 60 seconds;
    %all nodes dynamically adapt the rates.}
    \label{fig:exp_hc_comp}
\end{figure}

% We first evaluate oCSMA with and without channel information and 802.11 DCF in a static
% setting where the AP and the clients remain stationary throughout the experiment with
% fixed but heterogeneous PHY rates. Specifically, we perform the experiment using three
% different capacities: 6, 18, and 48 Mb/s for three clients.

First, in the static scenario, Figure~\ref{fig:exp_hc_thruput_comp} shows
the average per-flow throughput. We observe that only O-DCF can attain
proportional fair rate allocation in an efficient manner, because of the
consideration of the different link rates in the choice of CWs and
transmission lengths, whereas the other protocols show severe inefficiency
that the flows with higher rates are significantly penalized by the flow
with lower rate. In addition to the long-term fairness study in the
static scenario, we examine the responsiveness of O-DCF to channel
variations in the mobile scenario.  The mobile node's movement affects
its channel conditions, as shown in
Figure~\ref{fig:exp_hc_auto_rate_trace}.  We trace the instantaneous
throughput of both fixed and mobile nodes in
Figure~\ref{fig:inst_thruput_trace_smet}, where we see that the
incorporation of the runtime PHY rate in O-DCF indeed helps in achieving
throughput efficiency instantaneously even in the auto-rate enabled
environment.

\subsection{Experiment in Random Topology}

\begin{figure}[t!]
   \centering
   \subfigure[Tested scenario]{
         \includegraphics*[width=0.47\columnwidth]{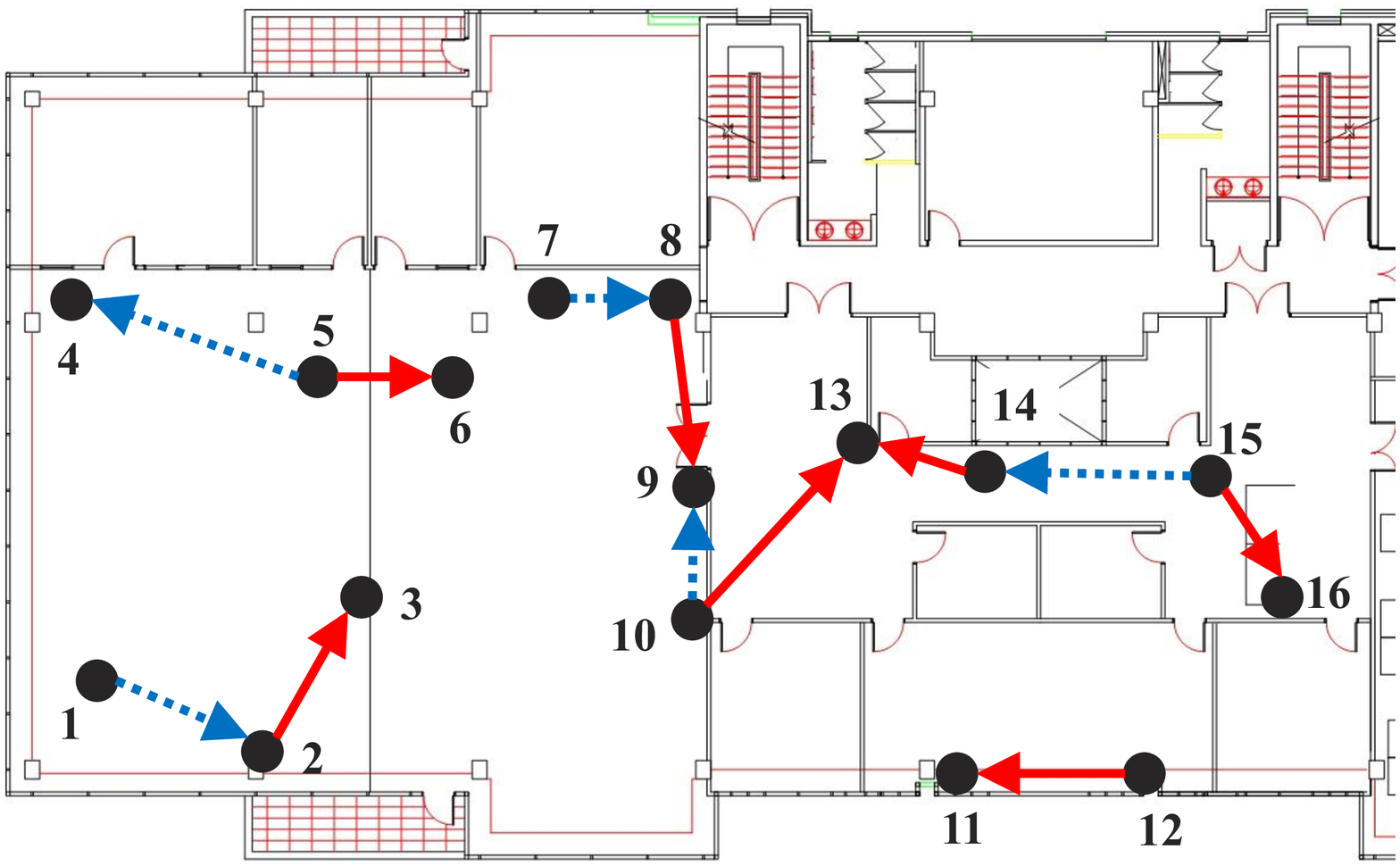}
         \label{fig:rand_topo}
   }\hspace{-0.1cm}
   \subfigure[Jain's fairness]{
         \includegraphics*[width=0.47\columnwidth]{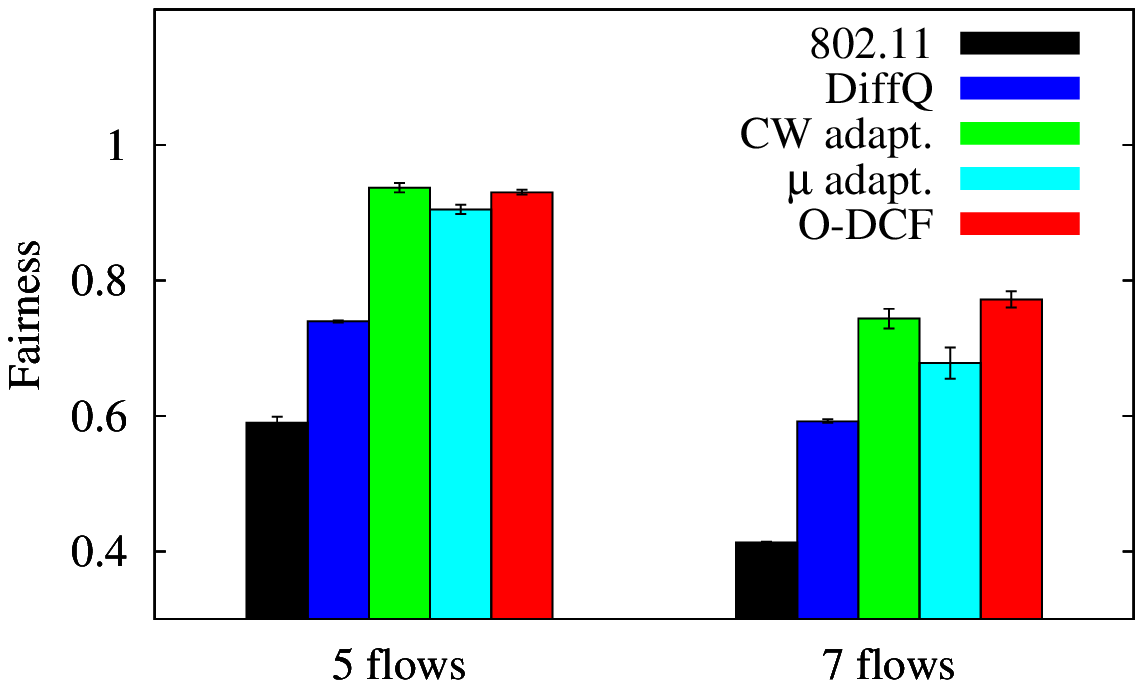}
         \label{fig:rand_fairness_comp}
   }
   \subfigure[Per-flow throughput of 5 flows]{
         \includegraphics*[width=0.46\columnwidth]{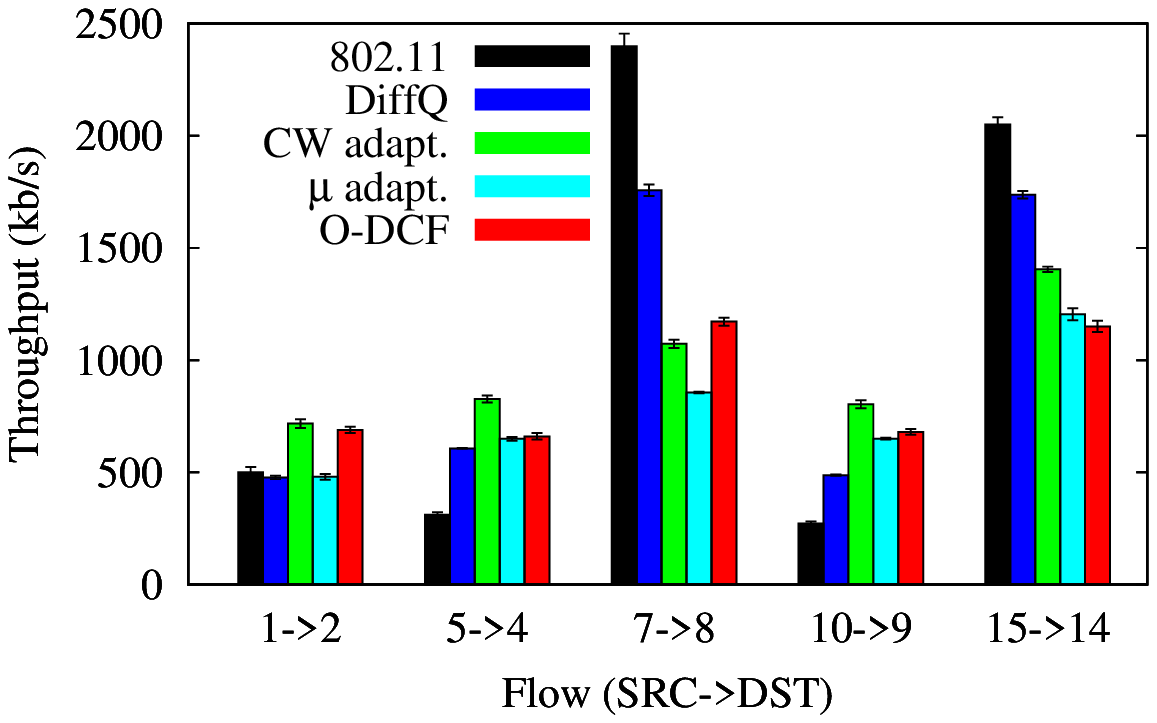}
         \label{fig:rand_5f_thruput}
   }
   \subfigure[Per-flow throughput of 7 flows]{
         \includegraphics*[width=0.46\columnwidth]{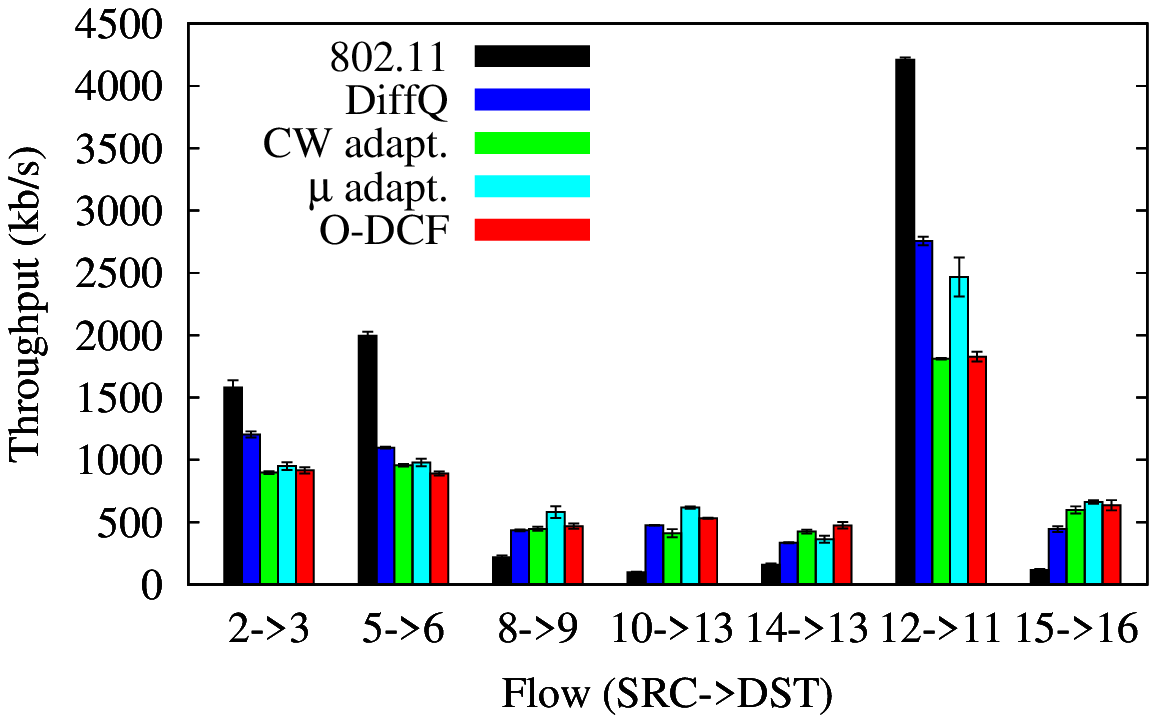}
         \label{fig:rand_7f_thruput}
   }
   \caption{Experiment. Tested topology and performance comparison; (a) 16 nodes denoted by triangles
   are distributed in the area of 40m x 20m; dotted (solid) arrows represent 5 (7) flows
   for the first (second) scenario. (b) Jain's fairness comparison among tested algorithms.
   (c)-(d) Per-flow throughput distributions.}
   \label{fig:rand_performance}
\end{figure}

We have so far focused on simple topologies representing their unique
features in terms of contention and sensing situation.  We now
perform more general experiments using a 16-node testbed topology, as
shown in Figure~\ref{fig:rand_topo}. We compare five algorithms in the
network with five and seven concurrent flows. Using
this random topology, we evaluate the impact of mixed problems, such as
hidden terminals and heavy contention scenarios including FIM scenario.
%We construct scenarios with a varying number of concurrent flows. The source and destination of each
%single-hop flow is chosen randomly. For a given number of flows, we experiment ten scenarios with
%different pairs of flows. The duration of each run is 60 seconds, in which the throughput of each
%flow is recorded at the end.
%The performance metrics we choose are Jain's fairness index \cite{JAIN84}, average per-flow throughput,
%and the summation of logarithmic utility of all the flows that measures the degree of proportional fairness.

Figure~\ref{fig:rand_fairness_comp} compares Jain's fairness achieved by
all the algorithms for two scenarios.  We find that over all the scenarios,
O-DCF outperforms others in terms of fairness (up to 87.1\% over 802.11 DCF
and 30.3\% over DiffQ), while its sum utility is similar with others. The
fairness gain can be manifested in the distribution of per-flow throughput,
as shown in Figures~\ref{fig:rand_5f_thruput} and \ref{fig:rand_7f_thruput}.
%Even if oCSMA achieves a bit smaller aggregate throughput (yet, comparable network
%utility) than others, it grants more access chances to the less-served
%links due to the sigmoidal mapping of queue-access probability and
%longer transmission time to the links with many collisions for
%reflecting the reduced access probability from BEB operation.
O-DCF effectively prioritizes the flows with more contention degree (e.g.,
flow $10\to9$ forms {\em flow-in-the-middle} with flows $7\to8$ and $15\to14$)
and provides enough transmission chances to highly interfered flows (i.e., $8\to9$,
$10\to13$, and $14\to13$), compared to 802.11 DCF and DiffQ. %, which are starved by some hidden flow $12\to11$.
%and is robust to hidden terminals (e.g., flow $13\to9$ are starved by two hidden flows $7\to8$ and $10\to12$)
%due to BEB with large CW values so that the hidden nodes can succeed in their transmissions.
The experimental topology is somewhat limited in size due to physical structure
of our building, so that the topology is close to be fully-connected, leading to
a small performance gap between oCSMA and O-DCF.
On the other hand, 802.11 DCF yields severe throughput disparities of more than
40 times between flows $12\to11$ and $10\to13$ in the second scenario.
%very low throughput of some flows
%(e.g., flow $9\to8$ in the 5-flow scenario and flow $13\to9$ in the 7-flow scenario).
%In the first scenario, the flow $9\to8$ forms {\em flow-in-the-middle} topology with
%the remaining flows. In the second scenario, the flow $13\to9$ are interfered by many
%other flows including {\em hidden flows} $7\to8$.
Compared to 802.11 DCF, DiffQ performs fairly well in the sense that it prioritizes highly interfered flows. 
However, its access prioritization is heuristic, so there is
still room for improvement toward optimality where compared to O-DCF.
%produces the starvation of flow $13\to9$ in the second one.
%This is due to the fact that DiffQ is vulnerable to hidden terminals, as mentioned in \cite{ASSI09}.

%However, the experimental topology is somewhat limited in size due to the physical
%structure of our building, leading to almost connected topology except small
%combinations of the links and very high degree of contention in the overall
%scenarios. Thus, we evaluate the performance of oCSMA in larger-scale scenarios
%using simulations in the next section.

%For better readability of graphs, the throughput values from different flow indexes are arranged in an increasing order.
%We verify that oCSMA achieves much improved fairness performance (the improvement in Jain's fairness
%index is at least 57\% in low density to 135\% in high density), compared to 802.11 where the gap
%between the rich and the poor is getting bigger as the flow density increases. This is due to the facts
%that (i) oCSMA significantly reduces the number of collisions irrespective of flow density and (ii)
%oCSMA allows less served flows to transmit more packets in order to maximize utility based fairness.

%\begin{figure}[t!]
%   \centering
%         \includegraphics*[width=0.45\columnwidth]{figure/largescale.eps}
%   \caption{Tested floor plan: 15 nodes denoted by black triangles are distributed in a floor of a building whose space is about 40 m x 20 m.}
%   \label{fig:rand_topo}
%\end{figure}

%\begin{figure}[t!]
%   \centering
%   \subfigure[3 flows]{
%         \includegraphics*[width=0.45\columnwidth]{figure/rand_3f_cdf.eps}
%         \label{fig:rand_3f_cdf}
%   }
%   \subfigure[6 flows]{
%         \includegraphics*[width=0.45\columnwidth]{figure/rand_6f_cdf.eps}
%         \label{fig:rand_6f_cdf}
%   }
%   \subfigure[9 flows]{
%         \includegraphics*[width=0.45\columnwidth]{figure/rand_9f_cdf.eps}
%         \label{fig:rand_9f_cdf}
%   }
%   \subfigure[12 flows]{
%         \includegraphics*[width=0.45\columnwidth]{figure/rand_12f_cdf.eps}
%         \label{fig:rand_12f_cdf}
%   }
%   \caption{Experiment. Distribution of per-flow throughput between oCSMA, DiffQ,
%   and 802.11 DCF in the random topology of Fig. \ref{fig:rand_topo}.}
%   \label{fig:rand_cdf}
%\end{figure}

%%% Local Variables:
%%% mode: latex
%%% TeX-master: "main"
%%% End:

\subsection{Simulation in Large Topologies}

%\note{Question: why without p or mu adaptation in 6.5 and 6.6?}

\begin{figure}[t!]
   \centering
   \subfigure[Grid topology]{
         \includegraphics*[width=0.4\columnwidth]{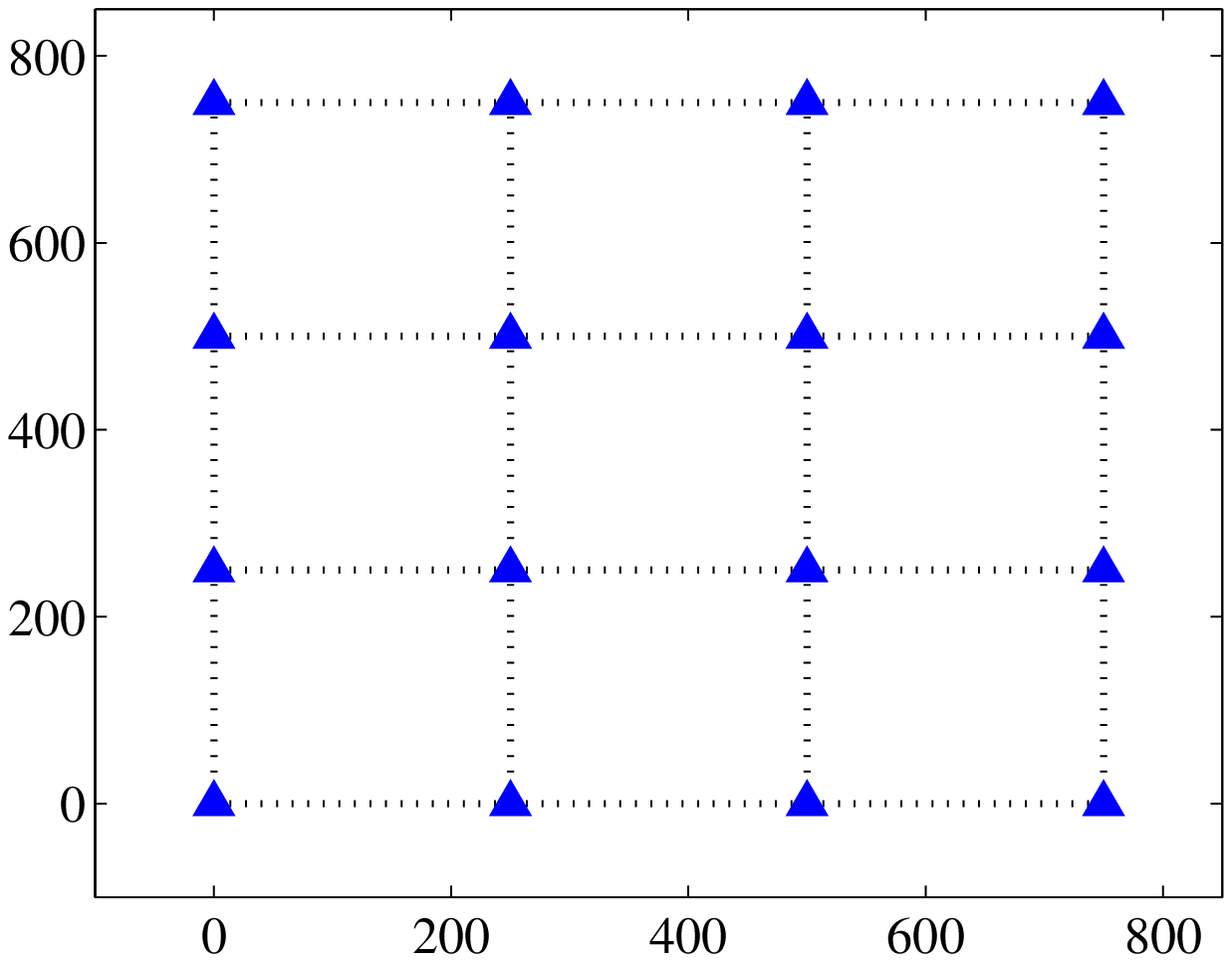}
         \label{fig:sim_grid_topo}
   } \hspace{0.2cm}
   \subfigure[Random topology]{
         \includegraphics*[width=0.4\columnwidth]{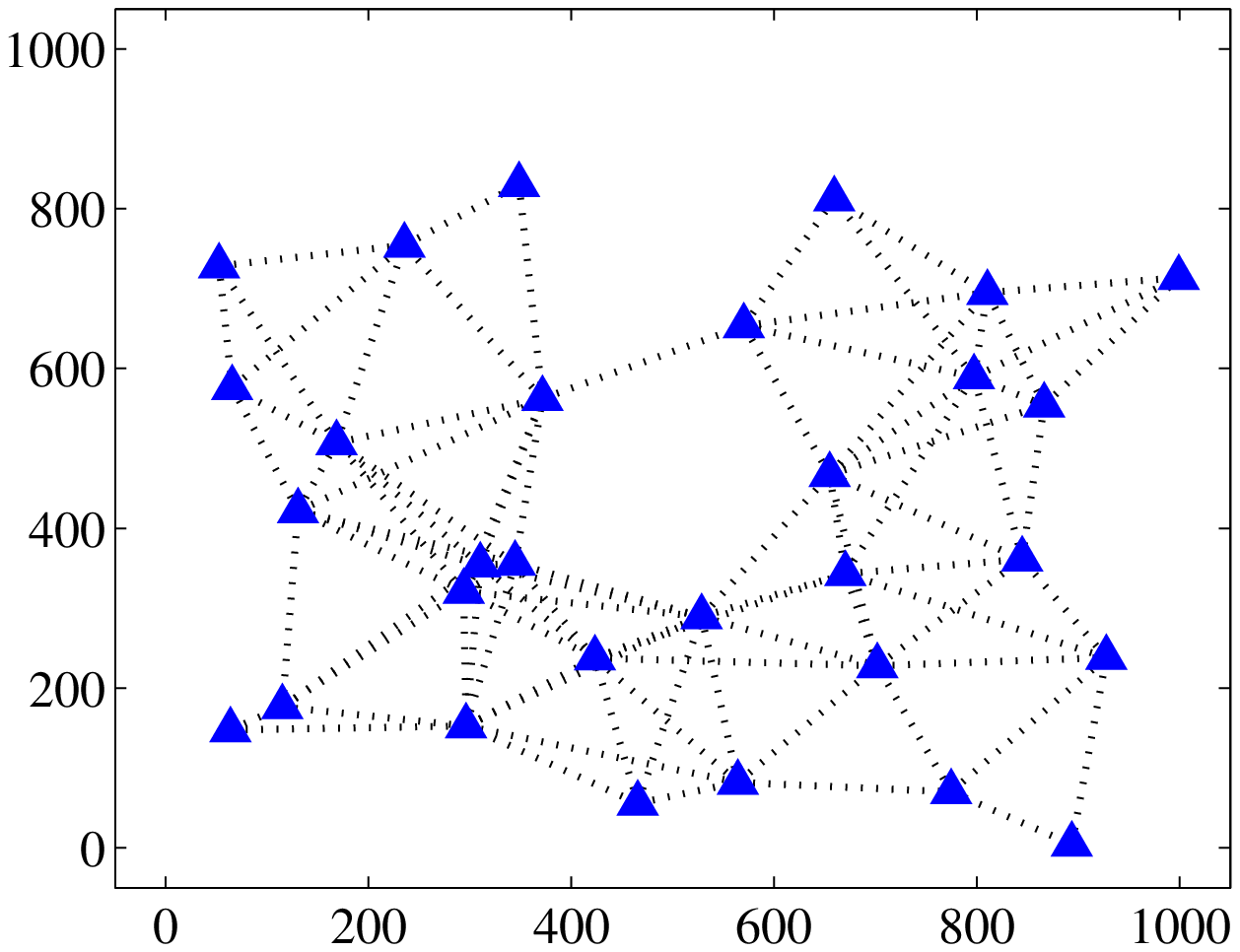}
         \label{fig:sim_rand_topo}
   }
   \subfigure[Jain's fairness]{
         \includegraphics*[width=0.48\columnwidth]{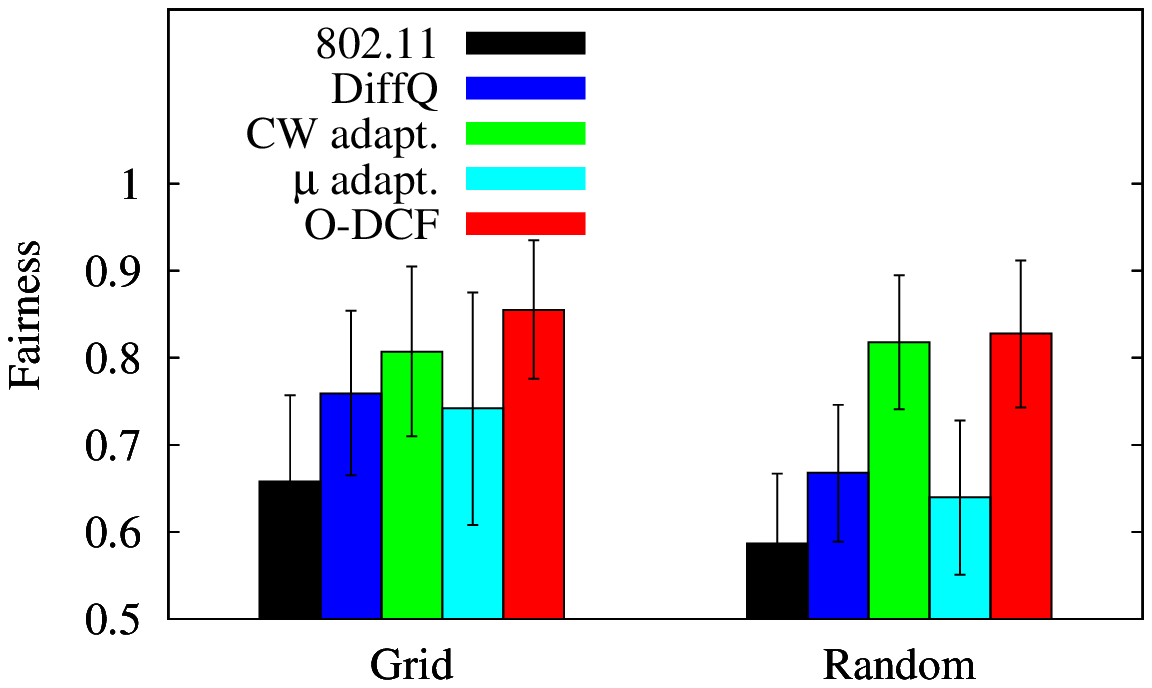}
         \label{fig:sim_ls_fairness_comp}
   }\hspace{-0.2cm}
   \subfigure[Sum of $\log$ utility]{
         \includegraphics*[width=0.48\columnwidth]{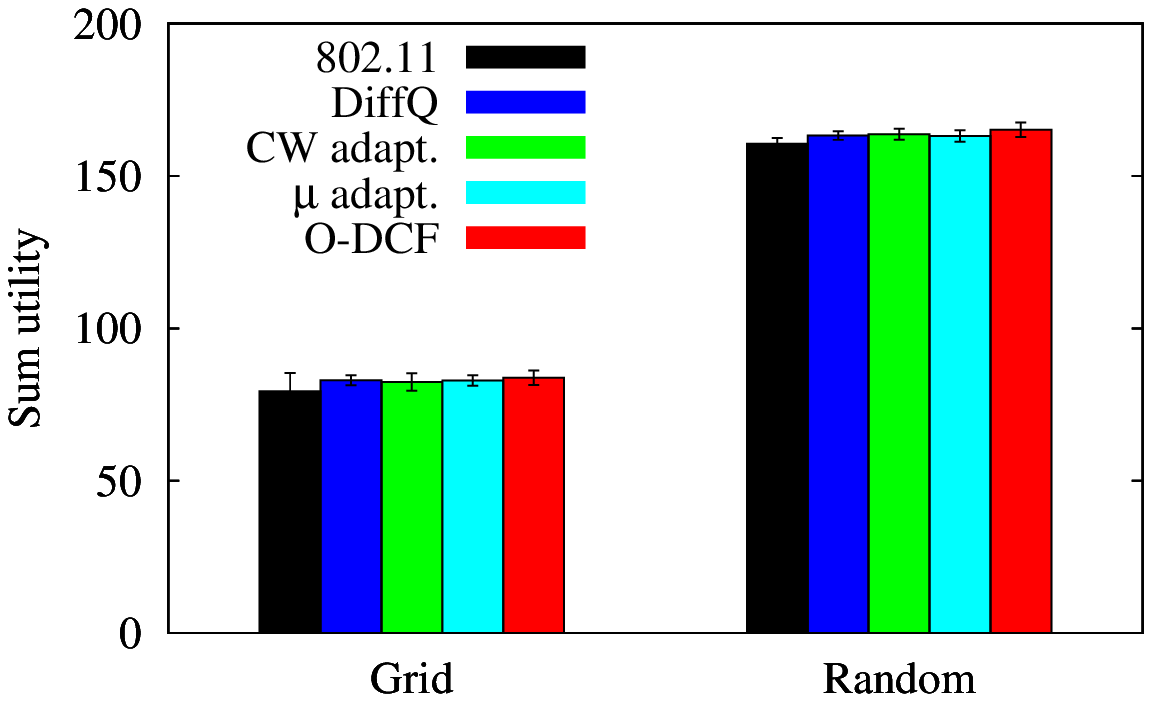}
         \label{fig:sim_ls_utility_comp}
   }
   \caption{Simulation. Large-scale topologies within 1000m $\times$ 1000m area
   and each node denoted by triangles has transmission and sensing ranges equal
   to 280m; (a) 16 nodes form a grid network; (b) 30 nodes are deployed randomly;
   (c)-(d) fairness and sum utility comparison among tested algorithms.}
   \label{fig:sim_ls_topo}
\end{figure}

Finally, we extend the evaluation of O-DCF to two large-scale scenarios in a
wider area. The first is a grid network where 16 nodes are apart from
each other with 250m, and the second is a random network where 30 nodes
are deployed randomly within an area of 1000m $\times$ 1000m. These
networks, depicted in Figures~\ref{fig:sim_grid_topo} and \ref{fig:sim_rand_topo},
contain a mixture of the previously discussed problematic scenarios,
such as HT, IA, and FIM, as well as highly interfering FC groups.
For a given number of flows, we construct 10 scenarios with different
random single-hop flows. In particular, we simulate 6 and 12 flows over
grid and random networks, respectively.
% , and allow the
% RTS/CTS mechanism.

Figure~\ref{fig:sim_ls_fairness_comp} compares the Jain's fairness
achieved by five algorithms. O-DCF enhances fairness up to
41.0\% (resp., 29.9\%) over 802.11 DCF and 24.0\% (resp., 12.6\%)
over DiffQ in the random (resp., grid) network.
%Such significant fairness improvement by O-DCF can be demonstrated
%by observing the distribution of per-flow throughput, as in
%Figures~\ref{fig:sim_grid_cdf} and \ref{fig:sim_rand_cdf}.
%\note{please explain more on the distribution figure. }
Compared to the previous experiments, we observe more throughput disparities
of contending flows under DiffQ and $\mu$ adaptation.
In most scenarios, randomly chosen flows likely have problematic relationships
such as HT and IA. This leads to lower fairness of DiffQ and $\mu$ adaptation
than O-DCF and CW adaptation, since they are vulnerable to imperfect sensing scenarios,
as examined in Section \ref{subsec:imperfect_exp}. Moreover, O-DCF experiences much
less collisions than CW adaptation, thus leading to better throughput.
%flows under DiffQ and 802.11 DCF, since the chosen flows obviously have
%problematic relationships such as HT, IA and heavy contention groups in
%most scenarios when simulating these topologies.
Finally, our O-DCF consistently shows a slightly better performance in terms of
sum utility (see Figure~\ref{fig:sim_ls_utility_comp}).

\section{Conclusion} \label{sec:conclusion}

We presented O-DCF so that 802.11 can work close to optimal in practice. 
The major design issues to make 802.11 DCF optimal include contention
window selection and transmission length control against network contention,
imperfect sensing, channel heterogeneity, packet capture without any message passing.
Moreover, our proposed O-DCF can be implemented with simple software modifications
on top of legacy 802.11 hardware. Through extensive simulations and experiments using
16-node wireless testbed, we have demonstrated that O-DCF performs similarly to that
in theory, and outperforms other MAC protocols, such as oCSMA, 802.11 DCF, and
DiffQ, under a wide range of scenarios.
Among the vast literature on 802.11/CSMA, O-DCF seems to be the first one that is validated 
through hardware experiments to be both near-optimal in utility maximization's efficiency and 
fairness metrics and 802.11 legacy-compatible.
%To the best of our knowledge, this is the first
%implementation-oriented effort in the CSMA based optimal scheduling.

\balance
{
%\small
\footnotesize
\bibliographystyle{habbrv}
\bibliography{main}
}
\end{document}